\documentclass[fleqn,usenatbib]{mnras}

\usepackage{newtxtext,newtxmath}
\usepackage[T1]{fontenc}
\usepackage{ae,aecompl}

\usepackage{graphicx}
\usepackage{amsmath}
\usepackage{amssymb}
\newcommand{\comment}[1]{}
\usepackage{gensymb}

\title[Resolved infrared imagery of Apep]{The extreme colliding-wind system Apep: resolved imagery of the central binary and dust plume in the infrared}

\author[Y. Han et al.]{Y.~Han,$^{1}$
P.~G.~Tuthill,$^{1}$\thanks{Email: peter.tuthill@sydney.edu.au}
R.~M.~Lau,$^{2}$
A.~Soulain,$^{1}$
J.~R.~Callingham,$^{3,4}$
P.~M.~Williams,$^{5}$
\newauthor
P.~A.~Crowther,$^{6}$
B.~J.~S.~Pope$^{7,8,9}$ and
B.~Marcote$^{10}$
\\
$^{1}$Sydney Institute for Astronomy (SIfA), School of Physics, The University of Sydney, NSW 2006, Australia\\
$^{2}$Institute of Space \& Astronautical Science, Japan Aerospace Exploration Agency, 3-1-1 Yoshinodai, Chuo-ku, Sagamihara, Kanagawa 252-5210, Japan\\
$^{3}$Leiden Observatory, Leiden University, PO Box 9513, 2300 RA, Leiden, The Netherlands\\
$^{4}$ASTRON, Netherlands Institute for Radio Astronomy, Oude Hoogeveensedijk 4, Dwingeloo, 7991 PD, The Netherlands\\
$^{5}$Institute for Astronomy, University of Edinburgh, Royal Observatory, Edinburgh EH9 3HJ, UK\\
$^{6}$Department of Physics \& Astronomy, University of Sheffield, Sheffield, S3 7RH, UK\\
$^{7}$Center for Cosmology and Particle Physics, Department of Physics, New York University, 726 Broadway, New York, NY 10003, USA\\
$^{8}$Center for Data Science, New York University, 60 5th Ave, New York, NY 10011, USA\\
$^{9}$NASA Sagan Fellow\\
$^{10}$Joint Institute for VLBI ERIC, Oude Hoogeveensedijk 4, 7991 PD Dwingeloo, The Netherlands\\
\\
}

\date{Accepted XXX. Received YYY; in original form ZZZ}

\pubyear{2020}

\begin{document}
\label{firstpage}
\pagerange{\pageref{firstpage}--\pageref{lastpage}}
\maketitle

\begin{abstract}
The recent discovery of a spectacular dust plume in the system 2XMM~J160050.7--514245 (referred to as ``Apep'') suggested a physical origin in a colliding-wind binary by way of the ``Pinwheel'' mechanism.
Observational data pointed to a hierarchical triple-star system, however several extreme and unexpected physical properties seem to defy the established physics of such objects.
Most notably, a stark discrepancy was found in the observed outflow speed of the gas as measured spectroscopically in the line-of-sight direction compared to the proper motion expansion of the dust in the sky plane.
This enigmatic behaviour arises at the wind base within the central Wolf-Rayet binary: a system that has so far remained spatially unresolved.
Here we present an updated proper motion study deriving the expansion speed of Apep's dust plume over a two-year baseline that is four times slower than the spectroscopic wind speed, confirming and strengthening the previous finding. 
We also present the results from high-angular-resolution near-infrared imaging studies of the heart of the system, revealing a close binary with properties matching a Wolf-Rayet colliding-wind system. 
Based on these new observational constraints, an improved geometric model is presented yielding a close match to the data, constraining the orbital parameters of the Wolf-Rayet binary and lending further support to the anisotropic wind model.
\end{abstract}




\section{Introduction}

Wolf-Rayet (WR) stars embody the final stable phase of the most massive stars immediately before their evolution is terminated in a supernova explosion \citep{Carroll2007}. They are responsible for some of the most extreme and energetic phenomena in stellar physics, launching fast and dense stellar winds that power high mass-loss rates \citep{Crowther2007}. When found in binary systems with two hot wind-driving components, a colliding wind binary (CWB) is formed, which may produce observational signatures from the radio to X-rays. Among the wealth of rare and exotic phenomenology associated with CWBs, perhaps the most unexpected is the production of copious amounts of warm dust \citep{Williams1987, Zubko1998, Lau2020}, which can occur when one binary component is a carbon-rich WR star (WC) whose wind provides favourable chemistry for dust nucleation. However, considerable controversy surrounds the mechanism by which some of the most dust-hostile stellar environments known (high temperatures and harsh UV radiation fields) are able to harbour protected dust nurseries which require high densities and relatively low temperatures for nucleation \citep{Crowther2007, Cherchneff2015}. 

A major clue was presented with the discovery of the ``Pinwheel nebulae'' \citep{Tuthill1999, Tuthill2008} which pointed to dust formation along the wind-shock interface which is co-rotating with the binary orbit. Subsequent inflation as the dust is carried outwards in the spherically expanding wind results in structures that encode the small-scale orbital and shock physics in the orders-of-magnitude larger plume structures. Images of such systems can reveal the dynamics of the embedded colliding-wind binary at scales otherwise difficult to resolve. Given the rarity of Wolf-Rayet stars in the galaxy and the brevity of their existence, the discovery and monitoring of these systems provides valuable insight into the evolution of massive stars \citep{Hucht2001}. However, only a small handful of such dust plumes have been found to be large and bright enough for imaging with current technologies in the infrared \citep{Monnier2007}. 

The recent discovery of ``Apep'' (2XMM~J160050.7--514245) by \citet{Callingham2019} presented a colliding-wind binary with a spectacular spiral plume visible in the mid-infrared. Extending $\sim12^{\prime\prime}$ across, Apep's dust plume is two orders of magnitude larger than the stereotypical pinwheel nebulae WR~104 \citep{Tuthill2008, Soulain2018} and WR~98a \citep{Tuthill1999b} given their respective distances. Further analysis revealed the system to boast exceptional fluxes from the radio, infrared and X-rays, placing it among the brightest colliding-wind binaries known and a rival to famous extreme CWB systems such as $\eta$~Carinae.

Although near-infrared imagery resolved a double at the heart of the system consisting of a companion star north of a highly IR-luminous central component, the $0.7^{\prime\prime}$ separation of the visual double is an order of magnitude too wide to precipitate conditions required to create dust via the Pinwheel mechanism. Instead, the suggestion arose that the system must be a hierarchical triple star with a much closer Wolf-Rayet binary driving the dust formation lying hidden within the central component itself. 

The most enduring enigma to the underlying physics driving Apep was laid bare in the finding that the spectroscopically determined wind speed of $3400\pm200\,$km\ s$^{-1}$ was larger than the apparent expansion speed of the dust plume by a factor of six, assuming a distance of 2.4~kpc \citep{Callingham2019}. The model attempting to reconcile these conflicting data invoked a strongly anisotropic wind, never previously observed in CWBs, arising from the Wolf-Rayet system, suggested to be in the form of a slow equatorial flow and a fast polar wind. If real, such a wind speed anisotropy could be generated by at least one rapidly rotating Wolf-Rayet star in the system \citep{Callingham2019}. Such near-critical rotation, in turn, is the essential ingredient required to transmute the ultimate fate of the star as a supernova into a potential long-duration gamma-ray burst (LGRB) progenitor under the currently referenced collapsar model \citep{woosley93,Thompson1994,macfadyen99,Macfayden2000, Woosley2006}. If confirmed, this would constitute the first system of this type yet observed \citep{Callingham2019}. 

The detailed and clean geometry of the spiral nebula encodes properties of the inner binary, but is not well reproduced by existing models. Certain features suggest a stage in its history when dust production was turned on and off, while others seem difficult to fit with any level of precision using standard recipes for the expected morphology. Although a plausible illustrative geometric model has been successfully sketched out \citep{Callingham2019}, it fails to reproduce features accurately, with extensive searches of the available degrees of freedom offering little promise. This difficulty seems likely to point to additional model complexity. 

To address these open questions, we observed Apep with the European Southern Observatory's (ESO) Very Large Telescope (VLT) at near- and mid-infrared wavelengths. Observations with the NACO and VISIR instruments are described in Section~\ref{sec:obs}. The detection of the central binary is provided in Section~\ref{sec:NACO} and the proper motion of the dust plume is studied in Section~\ref{sec:VISIR}. These results are used to constrain a geometric model presented in Section~\ref{sec:dis}, and implications for the distance and wind speed of the system are discussed. The findings of this study are summarised in Section~\ref{sec:sum}.

\section{Observations}
\label{sec:obs}

\subsection{NACO observations}

Near-infrared imagery of the Apep system was obtained with the NACO instrument \citep{Lenzen2003, Rousset2003} on the VLT at Paranal Observatory, with observational dates and basic parameters listed in Table~\ref{obs}. The 2016 epoch consisted of more orthodox (filled pupil) imagery, immediately splitting the central region into a $0.7^{\prime\prime}$ binary denoted by \citet{Callingham2019} as the ``central engine'' and the ``northern companion''.

\begin{table*}
\begin{center}
\caption{NACO and VISIR observations of Apep used in this manuscript with the central wavelength ($\lambda_0$) and width ($\delta\lambda$) of each observing band listed. Narrow-band (NB) and intermediate-band (IB) filters were used in the 2016 epoch of near-infrared imagery. }
\label{obs}
\begin{tabular}{ l l l l l r}
 \hline
 Instrument & Filter Name & $\lambda_0$ / $\delta\lambda$ ($\mu$m) & Mask & Calibrator Star & Observing Date\\
\hline
 NACO & \textit{IB\_2.24} & 2.24  / 0.06 & Full pupil & HD 144648 & 25, 28 Apr 2016 \\
 & \textit{NB\_3.74} & 3.740 / 0.02 & Full pupil & HD 144648 & 28 Apr 2016 \\
 & \textit{NB\_4.05} & 4.051 / 0.02 & Full pupil & HD 144648 & 28 Apr 2016 \\
 & \textit{J}  & 1.265 / 0.25 & 7-hole & [W71b] 113-03 & 20, 24 Mar 2019 \\
 & \textit{J}  & 1.265 / 0.25 & 9-hole & HD 142489 & 21 Mar 2019 \\
 & \textit{H}  & 1.66  / 0.33 & 7-hole & [W71b] 113-03 & 20, 24 Mar 2019 \\
 & \textit{H}  & 1.66  / 0.33 & 9-hole & HD 142489 & 21 Mar 2019 \\
 & \textit{Ks} & 2.18  / 0.35 & 7-hole & [W71b] 113-03 & 20, 24 Mar 2019 \\
 & \textit{Ks} & 2.18  / 0.35 & 9-hole & HD 142489 & 21 Mar 2019 \\
 & \textit{L'} & 3.80  / 0.62 & 9-hole & IRAS 15539-5219 & 22 Mar 2019 \\
 & \textit{M'} & 4.78  / 0.59 & 9-hole & IRAS 15539-5219 & 22 Mar 2019 \\
 VISIR & \textit{\textit{J8.9}} & 8.72 / 0.73 & Full pupil & & 13 Aug 2016\\
 & \textit{J11.7} & 11.52 / 0.85 & Full pupil & & 23 Jul 2016\\
 & \textit{\textit{J8.9}} & 8.72 / 0.73   & Full pupil & & 31 Jul 2017\\
 & \textit{\textit{J8.9}} & 8.72 / 0.73   & Full pupil & & 21 May 2018 \\
 & \textit{\textit{Q3}} & 19.50 / 0.40 & Full pupil & & 5 Jun 2018 \\
 \hline
\end{tabular}
\end{center}
\end{table*}

Because of the likely role of the central engine in driving the dust plume, a second epoch of imagery was motivated spanning the $J$, $H$, $Ks$, $L'$ and $M'$ bands and employing the 7- and 9-hole aperture masks \citep{Tuthill2010}: an observing methodology demonstrated to recover information up to and beyond the formal diffraction limit of the telescope \citep{Tuthill2000a, Tuthill2000b, Tuthill2006, Monnier2007}. 
These observations were executed in the final proposal period of NACO before it was fully decommissioned in October 2019. For each filter and mask combination, observations of Apep and a PSF reference star were interleaved for the purpose of calibrating the interferograms of Apep against the telescope-atmosphere transfer function. 

\subsubsection{Data reduction}
\label{sec_datared}

We reduced NACO imagery using software developed within our group (\href{https://github.com/bdawg/MaskingPipelineBN}{github.com/bdawg/MaskingPipelineBN}) which performed background sky subtraction, gain correction, iterative bad pixel interpolation, cosmic ray removal and interferogram centring and windowing. Bad frames with outlying flux distributions were removed to produce stacked sequences of calibrated image frames. Sample reduced images are displayed in Fig~\ref{fig:NACO}. Data acquired with the $M'$ filter were of poor quality -- the signals in individual frames were below the threshold needed to perform basic cleaning and were not processed further.


The next step performed by default in the pipeline is to centre and window the stellar image from the larger raw data frame, however the presence of the relatively close $0.7^{\prime\prime}$ binary made it problematic to extract separately isolated interferograms for both components. 
We developed two bespoke methods to resolve this issue. 
One method simply scaled the usual data window function to be very tight, retaining the component of interest and suppressing the remainder (including the unwanted binary component). 
The second applied a multiplicative window to suppress the unwanted component, after which the data could be handled as normal by the default package. 
In practice, these two were found to produce similar results and there were no cases where significant structures were found in one but not the other reduction method.
An exception to these successful strategies was encountered in $L$-band data in which diffraction spreads the light sufficiently so that the central engine and northern companion merge and cannot be readily separated.
The $L$-band data were therefore not used for model fitting.

In all cases, window functions consisted of a two-dimensional super-Gaussian:
\begin{equation}
    w(x, y) = e^{-\left( \frac{x^2+y^2}{2\sigma^2}\right)^P}
\end{equation}
\noindent where P was set to 5 for aggressive windowing to account for the proximity of the two components. 

As the spatial structures of the central component and the northern companion are both open to investigation, we performed all three possible calibration combinations: (a) calibrating the central component against the separate-in-time PSF reference star, (b) calibrating the central component against the northern companion and (c) calibrating the northern companion against the PSF reference star. 

Following established practice for masking interferometry data \citep{Tuthill2000a}, we sampled Fourier spectra for each windowed data cube (the sampling pattern depending on the mask and wavelength) which were normalised to yield complex visibility data. We accumulated robust observables comprising the squared visibility and the closure phase (argument of the bispectrum) over the data cube, with the statistical diversity yielding the uncertainties on the derived quantities. In order to extract high-resolution information from the full-pupil (non-masked) data, we performed an identical analysis to these sets of data with pseudo-sampling corresponding to that of an 18-hole mask.
Although full-frame speckle interferometry codes might offer superior processing for these data, past success with this strategy of pseudo-masking has delivered successful outcomes and allowed for direct comparison with all data passing through the same reduction pipeline.

We discarded badly calibrated source-calibrator pairs and outlying visibility data for individual baselines upon inspection of squared-visibility plots. 
Small statistical errors of the closure phase (estimated based on scatter across the cube) are likely an underestimate of systematic errors and can bias subsequent model fitting. We set closure phase uncertainties below the threshold of $2/3$ of the median uncertainty to this threshold value to avoid subsequent over-fitting to these data points.

\subsection{VISIR observations}
Mid-infrared imagery of Apep was acquired with the VISIR instrument on the VLT \citep{Lagage2004}. Observations taken in 2016 revealed the $\sim$\,12$^{\prime\prime}$ circumstellar structure in the form of an elaborately detailed spiral plume, while proper motions registered with 2017 data yielded a dust expansion speed of $570\pm70\,$km\ s$^{-1}$ at their distance of 2.4\,kpc \citep{Callingham2019}. 
Data from the 2017 epoch suffered from a minor issue in the construction of the observing sequence files: the chop/nod throw setting was too small, so that custom analysis code was needed to disentangle overlapped regions yielding uncontaminated imagery of the plume. 

Given the critical nature of the proper motion study in motivating novel physics, we requested a third epoch of VISIR imagery to eliminate any chance for systematic offsets due to the problems with the 2017 observing scripts.
This third epoch of VISIR data was taken in 2018, doubling the time-baseline available to validate and refine the previously established proper motion estimates. 
As a secondary goal, data were also taken with VISIR's \textit{Q3} filter to search for more distant, cooler dust.
A log of all epochs of mid-infrared data taken is also given in Table~\ref{obs}.

\begin{figure*}
\centering
\includegraphics[width=17cm]{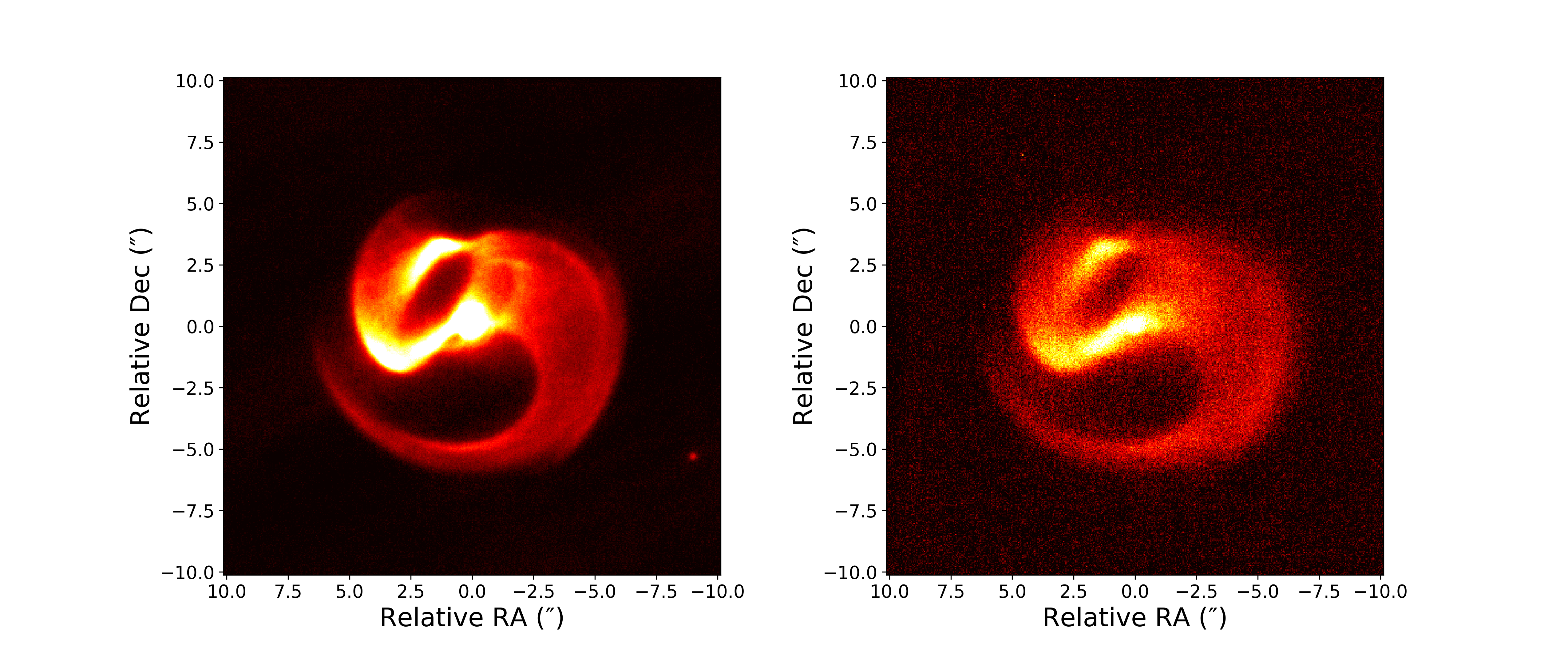}
\caption{Reduced 2018 epoch images taken with the \textit{J8.9} filter (left) and \textit{Q3} filter (right). The normalised pixel values were stretched from 0 to 0.03 in \textit{J8.9} and to 0.5 in \textit{Q3} on a linear scale. The faint background star used for plate scale calibration, 2MASS J16004953-5142506, is just visible to the south-west of the plume near the edge of the field in the \textit{J8.9} data.}
\label{fig:VISIR2018}
\end{figure*}

The reduction of VISIR data followed guidelines provided by the European Southern Observatory's VISIR manual \citep{VISIRmanual} to perform chopping and nodding subtractions. We calibrated the relative fluxes into units of Jy based on VISIR \href{https://www.eso.org/sci/facilities/paranal/instruments/visir/tools/zerop_cohen_Jy.txt}{standard} \comment{\footnote{\href{www.eso.org/sci/facilities/paranal/instruments/visir/tools/zerop_cohen_Jy.txt}{www.eso.org/sci/facilities/paranal/instruments/visir/tools/zerop\_cohen\_Jy.txt}}}observations of HD133550 (7.18 Jy at~J8.9 Filter), HD 178345 (8.21~Jy at B11.7 Filter) and HD145897 (1.84~Jy at Q3 Filter) taken on the same night as Apep observations with the corresponding filters. Given the high level of precision required to extract the modest proper-motion displacements of plume structures, we used a distant faint star, 2MASS J16004953-5142506, about $10^{\prime\prime}$ away from Apep and just bright enough to appear in stacked data to validate the consistency of the plate scale and orientation over the two-year interval. 
This yielded an upper limit on the magnitude of any deviations of plate scale of $0.4\%$.
Final stacked images obtained from the new 2018 data are given in Figure~\ref{fig:VISIR2018} and are discussed in Section~\ref{sec:VISIR}.

\section{Results: the central binary in the near-infrared}
\label{sec:NACO}
 
As the engine driving the colliding-wind plume and the origin of the presumed new physics underlying the discrepant wind speeds, the most critical observational challenge was to open a clear window onto the central binary.
This was accomplished by extracting spatial structures at the highest possible angular resolutions from the near-infrared data, however this process is not straightforward. 
The region has a number of complex elements (photospheres and dust) which vary with wavelength, are only partially resolved, and are not always well constrained by varying data quality over the observational record.
Competing models of increasing complexity for structure within the central component were confronted with these data and their plausibility discussed and evaluated in the sections below.

\subsection{Close binary star models}
\subsubsection{Fitting to closure phase data}

Starting with the simplest possible model, our binary star consists of two point sources and involves only three free parameters: binary separation, orientation and flux ratio. We fitted this model using the program \texttt{binary\_grid} (\href{https://github.com/bdawg/MaskingPipelineBN/blob/master/masking/util/binary_grid.pro}{github.com/bdawg/MaskingPipelineBN}) developed within our group for statistically robust identification of companions, particularly those at high contrast ratios.
The most likely model is that which globally minimises the $\chi^2$ error between the model and closure phase data, found by sampling a grid covering the parameter space and subsequently optimising with the Powell method \citep{Powell}. Given the Fourier coverage of our data, we estimated detection limits at confidence intervals of 99\% and 99.9\% using a Monte-Carlo method. The code injects a large number of noise-only (no binary) simulations into the algorithm to estimate the likelihood of false binary detections across the separation--contrast parameter space, which are used to define the corresponding confidence thresholds.

The best-fit solutions for the central binary largely (with one exception) calibrated against the PSF reference star (case (a) in Section~\ref{sec_datared}) are shown in Table~\ref{BinaryGrid}.
The most certain detections produced by \texttt{binary\_grid} exceeded 99.9\% confidence (for the \textit{IB\_2.24} filter), firmly establishing the reality of a close binary star within the central engine. 
Furthermore, recovered binary parameters from Table~\ref{BinaryGrid} are largely consistent across epochs and across filters despite some examples of more scattered values.
For example, visual inspection of the 2019 epoch revealed that the $J$-band data were of compromised quality.
Taken as a whole, the near-infrared data affirm the existence of a binary with the fainter component $\sim$47\,mas ($K$-band) to the West. 

\comment{
\begin{table*}
\begin{center}
\caption{Parameters of the central binary model comprised of two point sources fit to closure phase data only. A star (*) indicates a detection made with data calibrated against the northern companion. All other detections were made with data calibrated against the calibrator star.}
\label{BinaryGrid}
\medskip
\begin{tabular}{ l l l l l l }
 \hline
 Filter Name & Mask & Separation (mas) & Orientation ($^{\circ}$) & Contrast ratio & $\chi^2$\\
 \hline
 \textit{J}	 & 7-hole & 73.5 $\pm$ 4.6 & 241.8  $\pm$ 3.5 & 3.1 $\pm$ 1.4 & 110.4\\
 \textit{J}	 & 9-hole & 69.8 $\pm$ 22.9 & 28.0  $\pm$ 7.6 & 4.0 $\pm$ 2.2 & 20.0\\
 \textit{H}* & 7-hole & 29.0 $\pm$ 11.3 & 281.7 $\pm$ 9.5 & 4.0 $\pm$ 4.9 & 49.7\\
 \textit{H}  & 9-hole & 39.3 $\pm$ 5.7 & 277.6  $\pm$ 5.1 & 7.6 $\pm$ 1.5 & 49.7\\
 \textit{Ks} & 7-hole & 44.4 $\pm$ 9.4 & 274.0  $\pm$ 7.0 & 4.1 $\pm$ 1.6 & 77.2\\
 \textit{Ks} & 9-hole & 50.6 $\pm$ 5.3 & 279.2  $\pm$ 3.9 & 6.5 $\pm$ 1.0 & 49.9\\
 \textit{IB\_2.24} (\textit{K}) & Full pupil & 47.4 $\pm$ 2.9	& 273.8 $\pm$ 1.7 & 6.8 $\pm$ 0.7 & 309.8\\
 \hline 
\end{tabular}
\end{center}
\end{table*}
}

\begin{table*}
\begin{center}
\caption{Parameters of the central binary model comprised of two point sources fit to closure phase data only. A star (*) indicates a detection made with data calibrated against the northern companion. All other detections were made with data calibrated against the calibrator star.}
\label{BinaryGrid}
\medskip
\begin{tabular}{ l l l l l }
 \hline
 Filter Name & Mask & Separation (mas) & Orientation ($^{\circ}$) & Contrast ratio \\
 \hline
 \textit{J}	 & 7-hole & 73.5 $\pm$ 4.6 & 241.8  $\pm$ 3.5 & 3.1 $\pm$ 1.4 \\
 \textit{J}	 & 9-hole & 69.8 $\pm$ 22.9 & 28.0  $\pm$ 7.6 & 4.0 $\pm$ 2.2 \\
 \textit{H}* & 7-hole & 29.0 $\pm$ 11.3 & 281.7 $\pm$ 9.5 & 4.0 $\pm$ 4.9 \\
 \textit{H}  & 9-hole & 39.3 $\pm$ 5.7 & 277.6  $\pm$ 5.1 & 7.6 $\pm$ 1.5 \\
 \textit{Ks} & 7-hole & 44.4 $\pm$ 9.4 & 274.0  $\pm$ 7.0 & 4.1 $\pm$ 1.6 \\
 \textit{Ks} & 9-hole & 50.6 $\pm$ 5.3 & 279.2  $\pm$ 3.9 & 6.5 $\pm$ 1.0 \\
 \textit{IB\_2.24} (\textit{K}) & Full pupil & 47.4 $\pm$ 2.9	& 273.8 $\pm$ 1.7 & 6.8 $\pm$ 0.7 \\
 \hline 
\end{tabular}
\end{center}
\end{table*}

The alternate strategy ((b) in Section~\ref{sec_datared}) of calibrating the central binary against the northern companion yielded consistent results in the $H$ and $K$ bands to those shown in Table~\ref{BinaryGrid}.
Outcomes were also found to be robust under different choices of the windowing strategy to separate the $0.7^{\prime\prime}$ binary (also discussed in Section~\ref{sec_datared}).
Both findings further consolidate confidence in the detection made in this study. 

On the other hand, calibrating the northern companion against the PSF reference star ((c) in Section~\ref{sec_datared}) yielded inconsistent, low-confidence binary parameters.
This argues for the northern companion being an isolated single star at the scales of contrast and spatial resolution probed by this study.

\subsection{Models including embedded dust}
\subsubsection{Fitting to visibility data -- 1-component model}

While previous fits relied solely on closure phase data \citep[as is customary in companion detection with non-redundant masking,][]{Kraus2017}, the squared visibilities show a drop-off towards longer baselines at all position angles, requiring the existence of resolved circumstellar matter. The most natural scenario is local dust in the immediate vicinity of the binary. 
In this and the subsections that follow, emphasis in the modeling is given to the $K$-band data for which the quality is significantly higher resulting in more tightly constrained fits, and for which thermal flux from the dust becomes more pronounced.

We note in passing that the addition of dust also considerably complicates interpretation of the nature of the binary. 
The apparent fluxes may be influenced by local non-uniform opacity and/or thermal re-radiation from dust, so that the contrast ratios presented in the previous section may not correspond directly to the two stars.
This might permit, for example, a more nearly equal binary in luminosity as anticipated for a WR+WR \citep{Callingham2019,Callingham2020} despite the contrast data in Table~\ref{BinaryGrid}. 

In order to account for circumstellar material radiating in the near-infrared, we invoked models including spatially-resolved components, the simplest being a circular Gaussian flux distribution. Fits obtained with this model provided an estimate of the overall dimensions of the resolved component and are tabulated in column (a) of Table~\ref{table:gaussian}. The $J$-band data were not used for visibility-only fitting due to poorer phase stability delivered by AO at shorter wavelengths. Fitting to the visibilities in the $H$ and $K$ bands with a simple Gaussian model yielded a full-width at half maximum (FWHM) of approximately $36\pm4$\,mas, so that the spatial extent of the dust is comparable to the separation of the binary resolved in the previous section. 

\begin{table*}
\begin{center}
\caption{FWHM of the Gaussian component by (a) fitting a Gaussian source distribution to the visibility data and (b) fitting a Gaussian and two point source distribution to both visibility and closure phase data. }
\label{table:gaussian}
\medskip
\begin{tabular}{ l l c c }
 \hline
 Filter Name & Mask & (a) FWHM (mas) & (b) 3-component FWHM (mas) \\
 \hline
 \textit{J}	& 7-hole & No fit   & 44.7 $\pm$ 1.9\\
 \textit{J}	& 9-hole & No fit   & 48.7 $\pm$ 2.6\\
 \textit{H}	& 7-hole & 36.5 $\pm$ 3.7 & 30.0 $\pm$ 1.6\\
 \textit{H}	& 9-hole & 36.0 $\pm$ 3.7 & No fit\\
 \textit{Ks}	& 7-hole & 40.0 $\pm$ 4.0 & 29.5 $\pm$ 1.8\\
 \textit{Ks}	& 9-hole & 32.6 $\pm$ 3.3 & 27.0 $\pm$ 1.4\\
 \textit{IB\_2.24} ($K$) & Full pupil & 33.6 $\pm$ 3.4 & 26.4 $\pm$ 0.7\\
 \hline 
\end{tabular}
\end{center}
\end{table*}

\subsubsection{Fitting to visibility data -- 2-component models}
\label{sec:dust_flux}

In order to gradually add complexity to our model, a useful intermediate step is the combination of a resolved Gaussian dust component with an unresolved, embedded photosphere centred at the peak.
Confronting such a model to the data generally resulted in fits where the majority of the flux arose from the dust.
For example, the highest SNR data (narrow $K$ band) yielded a 2-component model with a best-fit value of 73\% of the total flux attributed to the resolved Gaussian. 

To indicate the impact of this on the interpretation, we adjust our earlier $\sim$6:1 $K$-band binary flux ratio now recognising the brighter component flux should be split two ways: a stellar photosphere and resolved dust shell. Performing this decomposition, we find that the embedded point sources are left with an approximately equal flux ratio ( $\sim$14\% of the total flux in each): a result that can be more neatly brought into accord with expectations from spectroscopic evidence \citep{Callingham2019,Callingham2020}. 

\subsubsection{Fitting to visibility and closure phase data -- 2-component models}

With an estimate of the relative contribution from the resolved component, we experimented with more complex multi-component models built with the interferometric data fitting program, \texttt{LITpro} \citep{Chesneau2009}. We replicated the model consisting of two simple point sources using \texttt{LITpro}'s gradient descent algorithm, now fitting to both the visibility and closure phase simultaneously in the $H$ and $K$ bands. This work successfully reproduced the binary detections presented in Table~\ref{BinaryGrid}. 

Simultaneous fitting to visibility and closure phase data was repeated with the 9-hole $J$-band data (previously flagged as of marginal quality). 
Given best estimate starting parameters, the algorithm yielded a roughly equal binary with a flux ratio of $1.0\pm0.4$. This appears to agree with the local dust hypothesis, a consequence of which is that the dominant contribution to the luminosity should shift from dust to photospheres as the observing band moves from $H$/$K$ to $J$ band. Indeed, although the poor data quality resulted in unconstrained fits with the simpler Gaussian source model, the drop in the squared visibilities towards longer baselines in the $J$ band appears much less pronounced than in the $H$ and $K$ bands, suggesting a much reduced dust fraction and a more unresolved morphology. 

\subsubsection{Challenges with higher degrees of freedom}

Physical expectations imply a 3-component model with two embedded point sources and a third resolved dust component (approximated here as a circular Gaussian, but it may also exhibit spatial structure). 
A model in which all 3 components can vary in flux, position and size (for the dust) introduces at least 7 degrees of freedom; parameters are also expected to vary with observing band and (possibly) over time.

Given limited and variable data quality, experimentation with models boasting such significant numbers of floating parameters were generally found to diverge or yield unreliable outcomes.
With a little perspective, this is hardly surprising.
All structures of interest are at or below the commonly-stated diffraction limit of the instrument and are modelled with highly covariant parameters.

To mitigate this problem, a conservative approach was taken, permitting only the minimum free model parameters and strictly delimiting possible options based on astrophysical plausibility and outcomes of earlier fits with simpler models.
Reliance on underlying physics eases the problem of requiring sufficient data to uniquely constrain the model.

\subsubsection{Fitting to visibility and closure phase data -- multi-component models}

To constrain our physically-motivated 3-component (2 stars plus dust shell) model, we begin by adopting the 47\,mas binary star separation and 274$^{\circ}$ orientation parameters from the $K$-band fit in Table~\ref{BinaryGrid}.
These 2-component closure-phase-only fits should be only weakly influenced by the dust, and largely constant with wavelength and epoch.
If we further adopt the idea that we are most likely dealing with a near-equal WR-WR binary \citep[flux ratio $\sim$1;][]{Callingham2020}, then we may use the 14\,:\,14\,:\,72\% flux partition established in Section~\ref{sec:dust_flux}.
Having established reasonable parameters of constraint for our model, the degree of freedom that remains is the spatial extent of the local infrared-bright hot dust: a quantity that would otherwise be entangled with ambiguities of the correct division of flux between dust and photosphere, and also with the true geometry of the binary. Fitting only to the FWHM of the Gaussian, the results of the $\chi^2$ minimisation are shown in Column~(b) of Table~\ref{table:gaussian}. The $H$-band 9-hole data could not be fitted by \texttt{LITpro} since the algorithm was unable to converge for poorly-calibrated visibility data in this set. 

Compared to the single component Gaussian model fit to visibility data only, the FWHM of the Gaussian in the 3-component model is slightly lower, which is likely due to the added luminous point source displaced from the centre of the source distribution. The apparent increase in Gaussian dust FWHM in the $J$ band is not interpreted as the physical extent of the dust, but is likely due to the assumed flux partition beginning to break down at short wavelengths.

\subsubsection{Best near-infrared 3-component model}
\label{sec:nir-bin}

In summary, the complex models discussed above demonstrate the likely existence of local dust extending over a spatial scale comparable to the binary separation. Synthesising information from model fitting and constraints based on basic stellar astronomy, we arrive at a plausible best-case model derived predominantly from $K$-band data to comprise an approximately equal binary separated by $47\pm6$\,mas embedded in a dust shell fit by a $35\pm7$\,mas-FWHM Gaussian. The position angle of the binary was determined to be $274\pm2^{\circ}$ in Apr 2016 and $278\pm3^{\circ}$ in Mar 2019. The binary photospheres rise to dominate the total flux in $J$ band, while conversely further in the infrared at $L$ band and beyond the dust rapidly becomes the dominant contribution, however declining data quality unfortunately presents challenges for the modelling at both.

\section{Results: the extended dust plume in the mid-infrared}
\label{sec:VISIR}

The second set of observational data listed in Table~\ref{obs} comprises a campaign of mid-infrared imaging with the VISIR instrument conducted to confirm the observed proper motions as well as to improve upon the derived parameters for the outflow. 

The displacement of the dust plume is modest even over the two-year time baseline now available, so that custom image processing procedures and algorithms were required for precise registration of the small (of order 1 pixel\ yr$^{-1}$) motions betraying wind-driven inflation of the circumstellar dust structure. We implemented an edge-detection algorithm (christened \texttt{ridge-crawler}) which has the capability locating edge-like features at the sub-pixel level. Images of Apep's dust plume were high-pass filtered to accentuate edge-like features and suppress regions of constant flux, effectively producing images of the skeleton of the spiral plume as shown in Figure~\ref{fig:ridges}. We then applied the algorithm, \texttt{ridge-crawler}, described in detail in Appendix~\ref{edgdet}, to extract the positions of the dust plume's edges, which were used to determine the expansion speed of the plume.

\begin{figure}
    \centering
    \includegraphics[width=7cm]{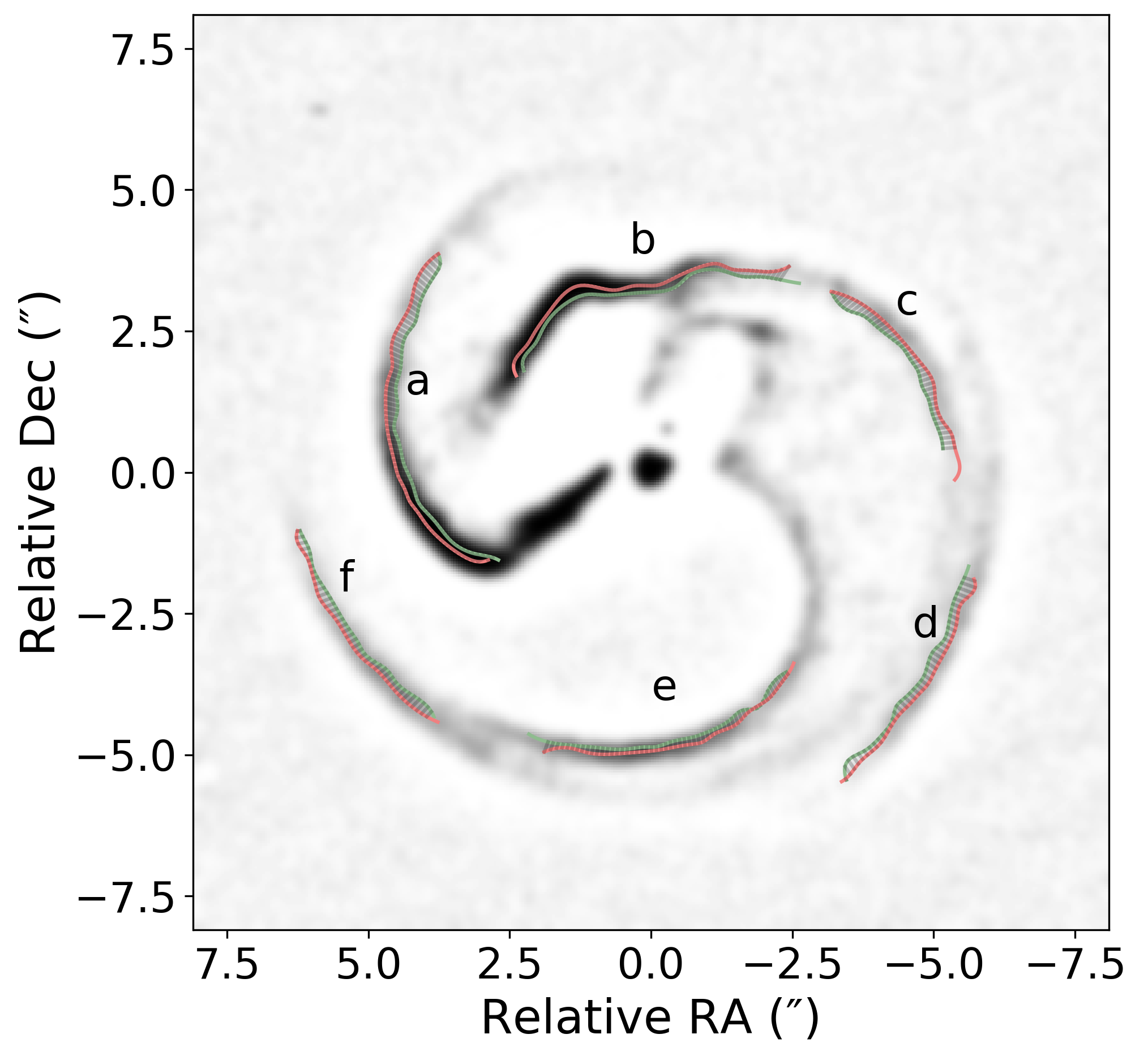}
    \caption{High-pass filtered 8.72 $\mu$m image of Apep labelled with the location of prominent edges detected by \texttt{ridge-crawler} at the 2016 (green) and 2018 (orange) epochs.}
    \label{fig:ridges}
\end{figure}

\subsection{The expansion rate of the dust plume}
The filtered skeleton of Apep highlights a number of prominent but disjoint edges as shown in Figure\,\ref{fig:ridges}, requiring each segment to be sampled and analysed individually. Table~\ref{speed} summarises the results returned by \texttt{ridge-crawler} calculated with a pixel scale of 45 mas/pixel \citep{VISIRmanual}.

\begin{table}
\begin{center}
\caption{Mean angular expansion speed of ridges $a$ to $f$ detected by \texttt{ridge-crawler} and identified in Fig. \ref{fig:ridges}. Unit: mas\ yr$^{-1}$. }
\label{speed}
\begin{tabular}{ c l l l } 
 \hline
 Ridge name & 2016 to 2017 & 2017 to 2018 & 2016 to 2018 \\
 \hline
 \textit{ a} & 63 $\pm$ 16 & 134 $\pm$ 26 & 95 $\pm$ 9  \\
 \textit{ b} & 61 $\pm$ 8  & 76  $\pm$ 17 & 67 $\pm$ 8  \\
 \textit{ c} & 67 $\pm$ 13 & 112 $\pm$ 28 & 85 $\pm$ 13  \\
 \textit{ d} & 24 $\pm$ 13 & 138 $\pm$ 32 & 74 $\pm$ 13  \\
 \textit{ e} & 61 $\pm$ 17 & 39  $\pm$ 18 & 51 $\pm$ 9  \\
 \textit{ f} & 38 $\pm$ 16 & 101 $\pm$ 24 & 66 $\pm$ 9  \\
 \hline
 \textbf{Mean} & \textbf{52} $\pm$ \textbf{14} & \textbf{100} $\pm$ \textbf{24} & \textbf{73} $\pm$ \textbf{10} \\
 \hline
\end{tabular}
\end{center}
\end{table}

The first point to note is that the 2018 data confirm the dust to exhibit an anomalously low proper motion.
All values recovered in Table~\ref{speed}, even the most extreme, are a factor of several too low to reconcile the angular rate of inflation on the sky with the line of sight wind speed of $\sim3400\,$km\ s$^{-1}$ from spectroscopy at any plausible system distance, which was estimated to be $\sim$2.4\,kpc \citep{Callingham2019}.

At a greater level of detail, the values exhibit scatter within a single epoch and between epochs larger than errors, even accounting for the intrinsically small level of the signal (less than VISIR's diffraction limit of $\sim\ 0.2^{\prime\prime}$ in the \textit{J8.9} filter \citep{VISIRmanual}). Turning firstly to the latter, differences in mean expansion rate between the three combinations of epochs are unlikely from any real behaviour of the plume (such as an acceleration). 
We attribute these to an artefact introduced by the custom image processing procedures used to recover images from the 2017 epoch (see Section~\ref{sec:obs}).
Given the consistent high quality data for both the 2016 the 2018 epochs, the values derived from this combination are therefore considered most reliable. 

On the other hand, some real diversity of speeds from ridges $a$ to $f$ within any single epoch is to be expected as different edges show differing apparent expansion rates projected onto the plane of the sky.
While 2-D images of an optically thin 3-D structure do result in most of the prominent edges close to the sky-plane (the effect of ``limb brightening''), departures are significant enough to cause measurable diversity as described in Appendix~\ref{sec:modelspeed}. 

To systematically study such projection effects, we constructed a model of the expanding dust plume (discussed in Section~\ref{sec:geo}) accounting for its three-dimensional structure, and calibrated the apparent motion of the edge-like structures identified above against the true (physical) expansion speed. Details of this calibration procedure is described in Appendix~\ref{sec:modelspeed}. Our results suggest that ridges $a$, $c$, $d$ and $f$ (group I) expand at approximately the true expansion rate of the system, while ridges $b$ and $e$ (group II) suffer apparent slowing due to projection effects. We therefore estimate the true expansion speed of the dust plume using the mean speed of group I ridges between the 2016 and 2018 epochs, obtaining a final value of $80\pm11$\,mas\ yr$^{-1}$. 

This new estimate corresponds to a linear expansion speed of approximately $910\pm120\,$km\ s$^{-1}$ assuming the previously published distance of $2.4$\,kpc \citep{Callingham2019}. At this distance, the dust expansion speed is then approximately four times slower than the spectroscopically determined wind speed. With the inclusion of a two-year time baseline, this analysis has therefore confirmed a consistently slow dust expansion. Since no current model or observation of colliding-wind binaries can explain such a drastic discrepancy between the outflow speed derived from the proper motions and that from spectroscopy, this finding strengthens the argument for an alternative model that deviates from the traditional picture of colliding-wind binaries. One such model is discussed in Section~\ref{sec:ani}.

\subsection{Spectral energy distribution}
\label{sec:SED}

The multiple epochs of Apep imagery across a number of infrared spectral bands provide information on the spectral energy distribution (SED) of the system. Building on the SED presented by \citet{Callingham2019}, we implemented custom procedures to extract the SEDs of individual components within Apep using NACO and VISIR data, namely the central binary, northern companion and extended dust plume. 

With half of the NACO data volume targeting calibrator stars whose photometric properties are well-determined, we extracted spectral flux densities of the central binary and northern companion by calibrating pixel values of background-subtracted, windowed sub-images of either component against those of the the corresponding calibrator star. 
NACO imagery also placed an upper bound of the flux contribution from the extended dust plume of below 5\%, which is not visible at these wavelengths at the SNR of the instrument. 

\begin{figure*}
\centering
\includegraphics[width=14cm]{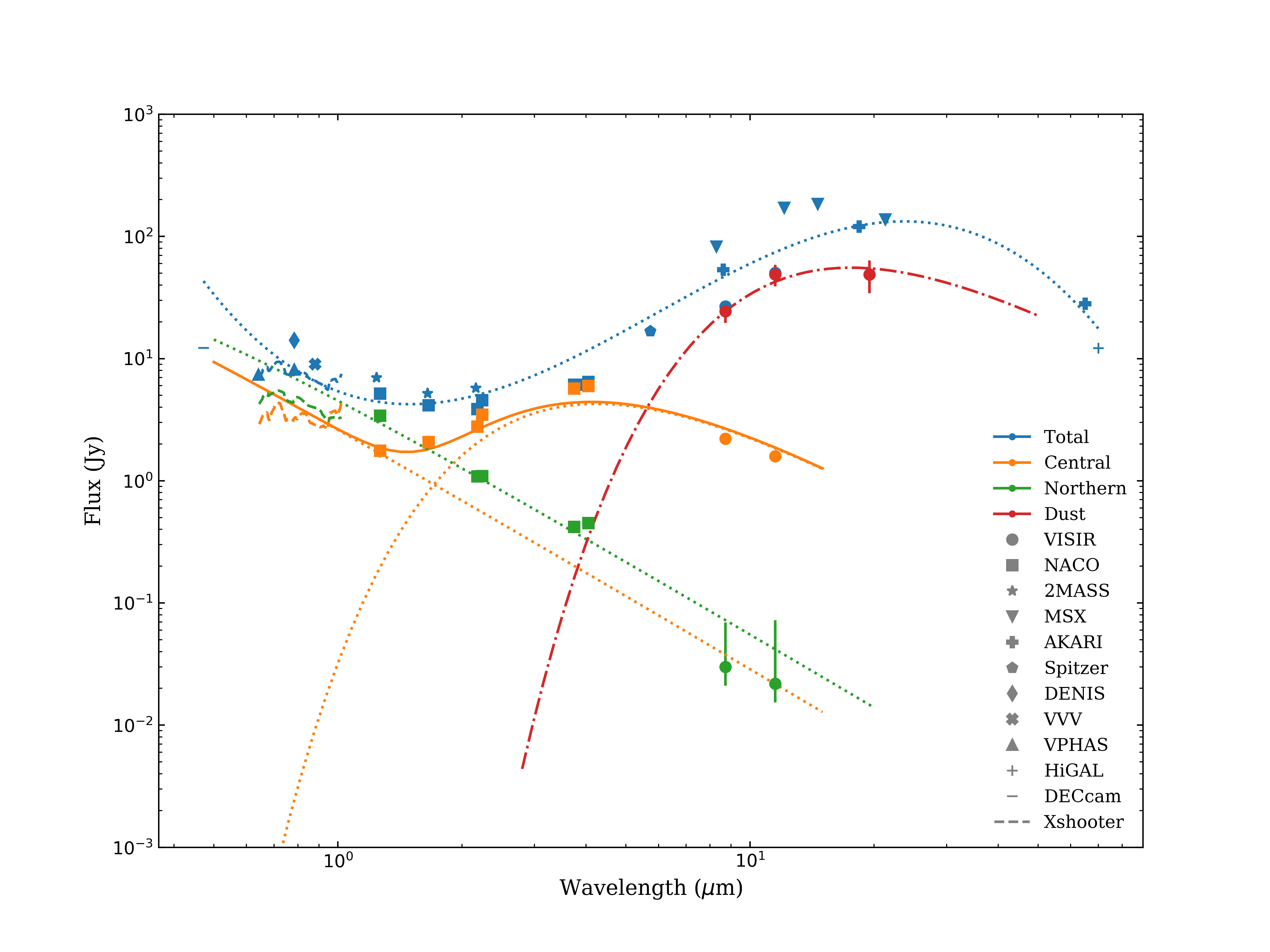}
\caption{The optical and infrared SED of Apep's components (central binary, northern companion and dust plume) dereddened with $A_v = 11.4$ \citep{Callingham2019, Mathis1990}. The SEDs of the Apep system and calibrator stars listed in Table~\ref{obs} are quoted from the \comment{Gaia (G, G\textsubscript{bp}, G\textsubscript{rp}) \citep{Gaia2016, Gaia2018}, }DECam (g) \citep{decam2018}, VPHAS+ (r, i) \citep{vphas2014}, DENIS (i) \citep{Denis1999}, VVV (Z) \citep{VVV2010}, 2MASS (J, H, Ks) \citep{2mass2006}, IRAC (5.8-micron) \citep{Spitzer2004}, MSX (A, C, D, E) \citep{MSX2001}, AKARI (S9W, L18W, N60) \citep{Akari2015} and Hi-GAL (70-micron) \citep{Higal2016} surveys and the calibrated XSHOOTER spectrum (moving average with 0.02-micron window) from \citet{Callingham2020}. The total flux of Apep was decomposed based on aperture photometry extracted from NACO and VISIR imagery. The northern companion is joined by a black-body spectrum at 34\,000\,K (dotted green line), the dust plume at 290\,K (dash-dotted red line), while the central binary (solid yellow line) can be modelled as the sum of two black bodies at 70\,000\,K and 1200\,K respectively (dotted orange lines).}
\label{fig:SED}
\end{figure*}

We used VISIR imagery to estimate the relative flux contributions from the central binary using PSF-fitting photometry. Since the northern companion lies on the shoulder of the much stronger mid-infrared peak from the central region, we isolated its relative flux by first subtracting the azimuthal average of the main peak. We attributed the remaining background-subtracted flux to the dust plume. 
The final SED is presented in Figure~\ref{fig:SED}.

We note that Apep exists in a complex region with extended background emissions \citep[as revealed by the Spitzer 8~$\mu$m image,][]{Spitzer2004}. The HiGAL 70-micron \citep{Higal2016}, AKARI N60 \citep{Akari2015} and MSX \citep{MSX2001} surveys had effective beam sizes larger than Apep's imaged dust plume, and so are vulnerable to contamination from background emission or cooler Apep emission from earlier episodes of dust production. We omitted WISE data \citep{wise2010} from the SED due to the low SNR despite their use of profile-fitted magnitudes to correct for effects of saturation.

The data appear to suggest variation in the mid-infrared flux over time. The MSX (c. 1996) and AKARI (c. 2006) values are both larger than the VISIR flux by a factor of a few across the mid-infrared. For example, the $N$-band flux decreases from MSX to AKARI ($\delta t \sim$~10~yr) and to VISIR ($\delta t \sim$~20~yr), which is consistent with the interpretation of an expanding and cooling dust plume. The measurements at shorter wavelengths were consistent with NACO fluxes, suggesting that the underlying flux heating the dust did not significantly vary over the interval covered by the observations. 
Future work is required to explore the mid-infrared variability of the system in more detail and evaluate its consistency with variations in the plume geometry.

\section{Discussion}
\label{sec:dis}

\subsection{Detection of Apep's central binary}
\label{sec:bin}

The results in Section~\ref{sec:NACO} establish that Apep's ``central engine'' \citep{Callingham2019} consists of a close binary with a secondary component approximately $47\pm6$\,mas ($K$-band mean) to the West of the primary component. This result rules out the possibility of the secondary component being a neutron star or a black hole -- which could conceivably drive a spiral plume by gravitational reflex motion, analogous to that witnessed in some mass-losing red giant systems such as AFGL 3068 \citep{Mauron2006} -- or that of a single-star dust production mechanism driving the prominent spiral nebula. This conclusion is supported by the findings of \citet{Callingham2020}, in which an analysis of the visible-infrared spectrum taken with XSHOOTER on the VLT found that the central engine consists of two WR stars of subtypes WC8 and WN4-6b respectively, identifying Apep as the first double classical Wolf-Rayet binary to be observationally confirmed. 

Furthermore, \citet{Marcote2020} were able to image the wind-collision region of the central binary at radio frequencies with very-long-baseline interferometry using the Australian Long Baseline Array (LBA) in Jul 2018. In addition to observing the colliding-wind shock front at the expected location of Apep's central binary, the authors derived a best-fit binary position angle of 277$^{\circ}$ based on the geometry of the shock front. The best-fit binary orientation angles of $274\pm2$ (Apr 2016) and $278\pm3$ (Mar 2019) that we derived in Section~\ref{sec:NACO} based on the NACO $H$/$K$-band data are in close agreement with the results of \citet{Marcote2020} based on the LBA data and confirm the colliding wind binary interpretation. 

Although the detection of a change in the binary position angle is only marginally significant given the errors, the direction of change is indicative of a counter-clockwise orbit is consistent with the expected direction given the geometry of the spiral plume and of the right magnitude, estimated to be $2\pm1^{\circ}$ over interval, given the expected binary period (Section~\ref{sec:geo}). 

The multi-component model fitting results also suggest the existence of local dust on a spatial scale comparable to the binary, which complicates the interpretation of the fitted flux ratios between the two stellar photospheres. The existence of a third component is supported by the SED (Figure~\ref{fig:SED}) of the central binary, in which there appears to be a large contribution from a component at a temperature significantly lower than that expected of stellar photospheres. 

Given the existence of local hot dust, the fitted contrast ratios of the double point source model presented in Table~\ref{BinaryGrid} is a reflection of the sum of the dust and primary stellar components' fluxes compared to the secondary stellar component's flux. The $K$-band absolute magnitudes of WC8 and WN4-6b stars are typically $-5.3\pm0.5$ and $-4.6\pm0.7$ respectively \citep{Rate2020}. Given the spectral types classified by \citet{Callingham2020} and the positional assignment of the two stars based on \citet{Marcote2020} (assuming the WN component launches a larger wind momentum), we expect the flux ratio of the two photospheres to be between 0.2 and 1.6. To achieve a contrast ratio of $6\pm2$ in the $K$-band, we require dust centred near the WC component to contribute approximately between 50\% and 80\% of the total flux. The estimate provided in Section~\ref{sec:dust_flux}, though indicative only, is consistent with this range expected from typical WR fluxes. 

The dust emission appears to diminish in flux towards $J$ band, though we interpret the best-fit values with caution due to the marginal data quality of this subset of data. The existence of dust around the WC8 star also appears to be supported by the observation that the C\,\textsc{iv} 2.08\,$\mu$m and C\,\textsc{iii} 2.11\,$\mu$m lines from the WC8 in the SINFONI spectrum of the central binary \citep{Callingham2019} appear to be more diluted than the He\,\textsc{ii} 2.189\,$\mu$m line primarily associated with the WN star.

\subsection{Dust properties}
Dust masses can be derived from the IR emission of Apep's spiral nebula. Assuming the IR emission is optically thin, the mass of the emitting dust grains of size $a$ and temperature $T_d$ can be estimated by

\begin{equation}
M_d=\frac{(4/3)\,a\,\rho_b\,F_\nu\,d^2}{Q_d(\nu,a )\,B_\nu(T_d)},
\label{eq:Mass}
\end{equation}

\noindent
where $\rho_b$ is the bulk density of the dust grains, $F_\nu$ is the measured flux, $d$ is the distance to Apep, $Q_d(\nu,a)$ is the dust emissivity model, and $B_\nu(T_d)$ is the Planck function at frequency $\nu$ and dust temperature $T_d$. The circumstellar material around WC stars are believed to be composed of amorphous carbon dust \citep{Cherchneff2000}. We therefore adopt a grain emissivity model consistent with amorphous carbon \citep{Zubko2004} and a bulk density of $\rho_b=2.2$ $\mathrm{gm}$ $\mathrm{cm}^{-3}$ (e.g. \citealt{Draine2007}). A grain size of $a=0.1$ $\mu$m is assumed for the emitting amorphous carbon. The dust temperature is derived from the measured IR fluxes assuming optically thin emission that can be approximated as $F_\nu\propto B_\nu(T_d)\nu^{\beta}$, where $\beta$ is the index of the emissivity power-law. A value of 1.2 is adopted for $\beta$, which is consistent with the mid-infrared emissivity index for amorphous carbon grains \citep{Zubko1996}. 

The total measured J8.9 and Q3 fluxes from Apep are $F_{J8.9} = 18.09\pm1.81$ Jy and $F_{Q3} = 38.75\pm7.75$ Jy, where $\sim8.4\%$ of the J8.9 flux arises from the central system. The emission from the central system in the Q3 filter is negligible ($\lesssim5\%$) compared to that of the overall nebula. Reddening by interstellar extinction is corrected using the ISM extinction law derived by Mathis (1990) and the estimated extinction towards Apep of $A_V=11.4$. The extinction correction factors for the J8.9 and Q3 fluxes are 1.48 and 1.26, respectively. From the dereddened fluxes, we derive an isothermal dust temperature of $T_d = 227^{+20}_{-15}$ K and dust mass (Eq.~\ref{eq:Mass}) of $M_d = 1.9^{+1.0}_{-0.7}\times10^{-5}$ M$_\odot$ in the spiral plume assuming a distance of 2.4~kpc towards Apep. Assuming this dust was formed in $\sim38$~yr, the dust production rate for Apep is $\dot{M}_d \sim5\times10^{-7}$ M$_\odot$ yr$^{-1}$.

\subsection{Geometric model}
\label{sec:geo}

In order to develop a physical model of Apep, we first construct a geometric model for the dust plume. 
\citet{Callingham2019} presented a plausible geometric spiral model based on the standard Pinwheel mechanism \citep{Tuthill2008}, matching the gross features in the data. 
However, on close comparison, the locations of edges and sub-structures provided a relatively poor match when overlaid. 
Despite fairly extensive exploration of the parameter space, varying the spiral-generative parameters did not offer significant improvement. 

Here we present a new modelling effort aimed at creating a geometric model systematically linking the spiral nebula of Apep to the underlying binary. 
We expand upon previous generative models, adding eccentricity as an extra degree of freedom to the binary orbital parameters. 
Details of model construction are described in Appendix~\ref{geomod}.

\subsubsection{Fitted model}

\begin{figure*}
\centering
\includegraphics[width=17cm]{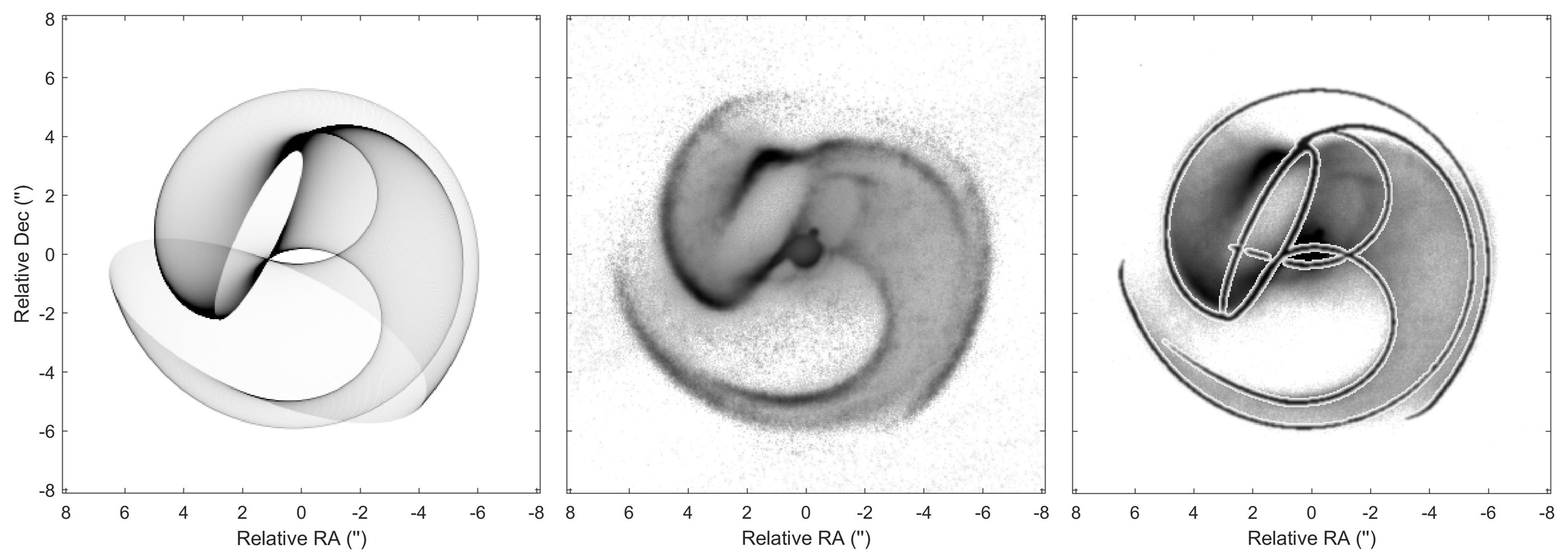}
\caption{Left: a geometric model of the dust plume of Apep. Middle: edge-enhanced image of Apep at $8.72\mu$m. Right: the outline of the geometric model overlaid on the data.}
\label{fig:Model}
\end{figure*}

\begin{table}
\begin{center}
\caption{Best-fit parameters of the geometric model of Apep assuming a dust expansion rate of 80$\pm$11~mas\,yr$^{-1}$ determined by the proper motion study. }
\label{geoparams}
\begin{tabular}{ l r } 
 \hline
 Parameter & Fitted value \\
 \hline
 Period ($P$)                                  & 125 $\pm$ 20 yr \\
 Eccentricity ($e$)                            & 0.7 $\pm$ 0.1 \\
 Cone full opening angle ($2\theta_{w}$)       & 125 $\pm$ 10 $^{\circ}$ \\
 Argument of periastron ($\omega$)             & 0 $\pm$ 15 $^{\circ}$ \\
 Inclination ($i$)                             & $\pm$ 25 $\pm$ 5 $^{\circ}$ \\
 Position angle of ascending node ($\Omega$)   & -88 $\pm$ 15 $^{\circ}$ \\
 Most recent year of periastron passage  ($T_0$)    & 1946 $\pm$ 20\\ 
 True anomaly at 2018 epoch ($\nu_\text{2019}$)     & -173 $\pm$ 15 $^{\circ}$\\
 True anomaly at dust turn-on ($\nu_\text{on}$)     & -114 $\pm$ 15 $^{\circ}$\\
 True anomaly at dust turn-off ($\nu_\text{off}$)   & 150 $\pm$ 15 $^{\circ}$\\
 \hline
\end{tabular}
\end{center}
\end{table}

A geometric model fitted to the mid-infrared data is presented in Fig~\ref{fig:Model}, in which the structural skeleton of the new model closely reproduces that of Apep's spiral plume extracted from the mid-infrared images. 

The best-fit parameters of the model are presented in Table~\ref{geoparams}, where the wind speed of $80\pm11$\,mas\ yr$^{-1}$ was constrained by the proper motion study in Section~\ref{sec:VISIR}. The model is also consistent with the near-infrared position angle of the binary discussed in Section~\ref{sec:bin}, which we used to constrain the true anomaly at the time of observation. 

Along with assumptions of Apep's distance (discussed in Section~\ref{sec:dist}) and near-infrared angular separation which together constrain the semi-major axis, the orbit of Apep's central binary is almost uniquely constrained. The remaining two-fold degeneracy arises from the sign of the inclination which the geometric model (based on an optically-thin surface) does not constrain. In either case, the low inclination of the binary implies a mostly polar view onto the spiral plume. 

\subsubsection{Eccentricity and episodic dust production}
Most notably, a relatively high orbital eccentricity of $0.7\pm0.1$ was required to fit the model of Apep. 
There exist examples of persistent dust producing WR stars \citep{Williams2015b} in which a prehistory of tidal interactions leading to orbital circularisation has been invoked to explain the observed lack of dependence of dust production on orbital phase \citep{Williams2019}. Circular orbits have been confirmed for CWBs such as WR~104 ($e<0.06$) \citep{Tuthill2008} and WR~98a \citep{Tuthill1999} based on the circular geometry of the Pinwheel spiral. 

On the other hand, the archetype episodic dust producer WR~140 with a very high orbital eccentricity ($e \simeq 0.9$) \citep{Fahed2011} does not form the classic spiral geometry, owing to the very brief duration of dust formation near periastron. 
Apep, like WR~140, is too wide for tidal circularisation, even during past giant phases. 
Although we have no historical record of variations in infrared flux (which would require time baselines of $\sim$125\,yr), the central binary of Apep seems a much closer analog to the WR~140 system than to persistent dust producers such as WR~104. 

Models for episodic dust producers such as WR~140 explain the periodicity of dust production in terms of the binary's orbital phase, with periastron approaches creating wind collisions strong enough to mediate dust nucleation \citep{Williams1990,Usov1991}. Our new model for Apep's plume suggests that the production of dust is indeed centred near periastron, with the central true anomaly of the dust-producing zone offset by $\sim$18$^{\circ}$ after periastron (or $\sim$~1~yr). The dust, which spans $\sim$264$^{\circ}$ over the orbit, was produced over a period of $\sim$38~yr between 1936 and 1974. 

In light of the geometric model, we can clearly identify the first and final dust rings of the spiral plume. The ratio of their radii ($\sim6.5^{\prime\prime}$ and $\sim12^{\prime\prime}$) are consistent with that between the number of years ago that they were produced at the time of observation. Assuming episodic dust production, our model also predicts the next cycle of dust production to begin at around 2061, with the next periastron passage occurring at around 2071. 

The distinct ring-like features corresponding to the turning on and off of dust production appears to support the threshold effect of dust production as seen in WR~140 when conditions of nucleation are briefly reached near periastron \citep{Williams2009}. Observationally, this threshold effect may be confirmed from the detection of more distant dust from a prior dust production cycle concentric to the prominent spiral plume of Apep. 
Expectations for the location of the previous coil of the plume are given in Figure~\ref{fig:modelpics}.
Future studies may benefit from instruments such as ALMA to search for distant dust structures expected from the threshold effect. 

Another observational confirmation of episodic dust production may come from the production of dust from a newer cycle close to the central binary. Based on the expected time frame of the upcoming episode of dust production, it is unlikely that hot, local dust as part of a new cycle of dust spiral via the Pinwheel mechanism is responsible for the bright mid-infrared peak at the central engine. Alternative dust production mechanisms, such as those leading to eclipses observed in WR~104 \citep{Williams2014}, may instead be responsible. 

Observations with instruments capable of delivering higher angular resolution (e.g. SPHERE or MATISSE on the VLT) may reveal the nature of the local dust and search for the potential upcoming cycle of colliding-wind mediated dust production.

\subsubsection{Opening angle}

We also note that the opening angle of $2\theta_{w, \text{IR}} = 125\pm10^{\circ}$ derived from this geometric model of Apep is smaller than the value of $2\theta_{w, \text{radio}} = 150\pm12^{\circ}$ derived by \citet{Marcote2020} based on the geometry of the radio shock front at the wind-collision region. In interpreting this apparent discrepancy, it is important to note that the two values, $2\theta_{w, \text{IR}}$ and $2\theta_{w, \text{radio}}$, may be reflective of the shock structure at different orbital phases of this eccentric binary. Specifically, $2\theta_{w, \text{IR}}$ is a measurement of the opening angle when the shock was radiative to allow for efficient cooling and hence dust production. Following dust production centred near periastron, the shock structure may have been gradually modulated as the binary moved further apart, potentially resulting in the shocked region flaring out from the contact discontinuity as the wind shock becomes adiabatic \citep{Pittard2018}. 

According to theoretical relationships derived by \citet{Gayley2009}, a radiative shock with opening angle $2\theta_{w, \text{IR}}$ derived in this study is consistent in terms of wind-momentum ratio, $\eta$, with an adiabatic shock with mixing having an opening angle $2\theta_{w, \text{radio}}$ derived by \citet{Marcote2020}. 
The radiative opening angle would then imply a wind momentum ratio of $\eta_{\text{IR}} = 0.21\pm0.07$ based on the relationship derived by \citet[eqn 9]{Gayley2009}, a value lower than $\eta_{\text{radio}} = 0.44\pm0.10$ \citep{Marcote2020} calculated using the adiabatic opening angle. 

The orbital modulation may hence explain the broadening of the shock region reflected in the $2\theta_{w, \text{radio}}$ measurement, and is possibly associated with the brightening of the 843-MHz radio flux reported by \citet{Callingham2019} which has so far lacked a satisfactory explanation. However, this suggestion requires detailed modelling in future studies to verify.

It is worth noting that the geometric model sketched above is intended to serve as an argument for self-consistency and plausibility of the underlying physical parameters. The results provide evidence that the morphology of Apep's spiral can indeed be produced via the Pinwheel mechanism in a colliding-wind binary, and that the underlying binary is likely in an eccentric orbit. The geometric model paves the way to a full radiative transfer model which may more accurately represent the variations in brightness across the spiral plume, adding a layer of physical modelling onto what appears to be a sound geometrical construct.

\subsection{Enclosed mass and distance to Apep}
\label{sec:dist}

Spectroscopic studies have estimated that the distance of Apep is $2.4^{+0.2}_{-0.5}$\,kpc \citep{Callingham2019}, which was more recently revised to $2.0^{+0.4}_{-0.3}$\,kpc \citep{Callingham2020}. If we were to assume the most recent distance estimate and the binary separation discussed in Section~\ref{sec:NACO}, we derive a physical WR binary separation of $95^{+19}_{-14}$\,AU at the time of NACO observations (corrected for projection). Together with constraints provided by the geometric model, the orbit of the central binary is uniquely constrained with a semi-major axis of $56^{+24}_{-17}$\,AU. Using a Keplerian model, we immediately estimate the enclosed mass of the WR binary to be $11^{+13}_{-6}$\,M$_{\odot}$. 

It is understood that the mean mass of WC8 stars is around 18\,M$_{\odot}$ \citep{Sander2019} although variations over a range of about 7\,M$_{\odot}$ \citep{Crowther2007} are possible.
WN stars have a much larger mass range with extreme cases up to 80\,M$_{\odot}$, but with the mean values typically under 20\,M$_{\odot}$. Given the spectral classification determined by \citet{Callingham2020}, we expect the enclosed binary mass of Apep to lie between 20 and 40\,M$_{\odot}$.

The mass estimate relying on a distance of $2.0^{+0.4}_{-0.3}$\,kpc is therefore lower than the most likely mass range. However, adopting the earlier $2.4^{+0.2}_{-0.5}$\,kpc \citep{Callingham2019} estimate would imply a semi-major axis of $67^{+20}_{-23}$\,AU and an enclosed mass of $19^{+12}_{-12}$\,M$_{\odot}$, which exhibits a larger overlap with the expected mass range. This analysis appears to argue for a distance towards the upper end of the spectroscopically determined range. We hereon adopt the $2.4^{+0.2}_{-0.5}$\,kpc value for subsequent calculations.

Confirmation of the $\sim2.4$\,kpc distance also argues for a common distance for the WR binary and the northern O-type supergiant in a hierarchical triple. 
Although chances of finding such field stars aligned within an arcsecond seems remote, the apparent lack of participation of the northern companion either in the radio or infrared has motivated debate on whether the stars are physically associated or not \citep{Callingham2020}.

\subsection{The wind speed dichotomy}
\label{sec:ani}

Adopting the $2.4^{+0.2}_{-0.5}$\,kpc system distance \citep{Callingham2020}, the linear expansion speed of the dust translates to $910\,^{+210}_{-290}\,$km\,s$^{-1}$. Comparing this value to the spectroscopic wind speed, this study supports the original finding that we are indeed observing a dust expansion significantly slower than either spectroscopic wind speed of the component WR stars: $3500\pm100\,$km\,s$^{-1}$ or $2100\pm200\,$km\,s$^{-1}$, launched by the two stellar components of the Wolf-Rayet binary \citep{Callingham2020}. 
The underlying conundrum presented by Apep in which the plane-of-sky (proper motion) expansion is starkly at odds with well-established line-of-sight (spectroscopic) windspeed is therefore reaffirmed.

Furthermore, we note that the Apep system seems unique in this behaviour: other Pinwheel systems including prototypes WR~140 \citep{Williams1990, Williams2009}, WR~104 \citep{Tuthill1999, Harries2004, Tuthill2008, Soulain2018} and WR~112 \citep{Lau2017, Lau2020submitted} (previously suspected as displaying ``stagnant shells'') all exhibit dust motions in accord with expectations from spectroscopy. 

\begin{figure*}
\centering
\includegraphics[width=15cm]{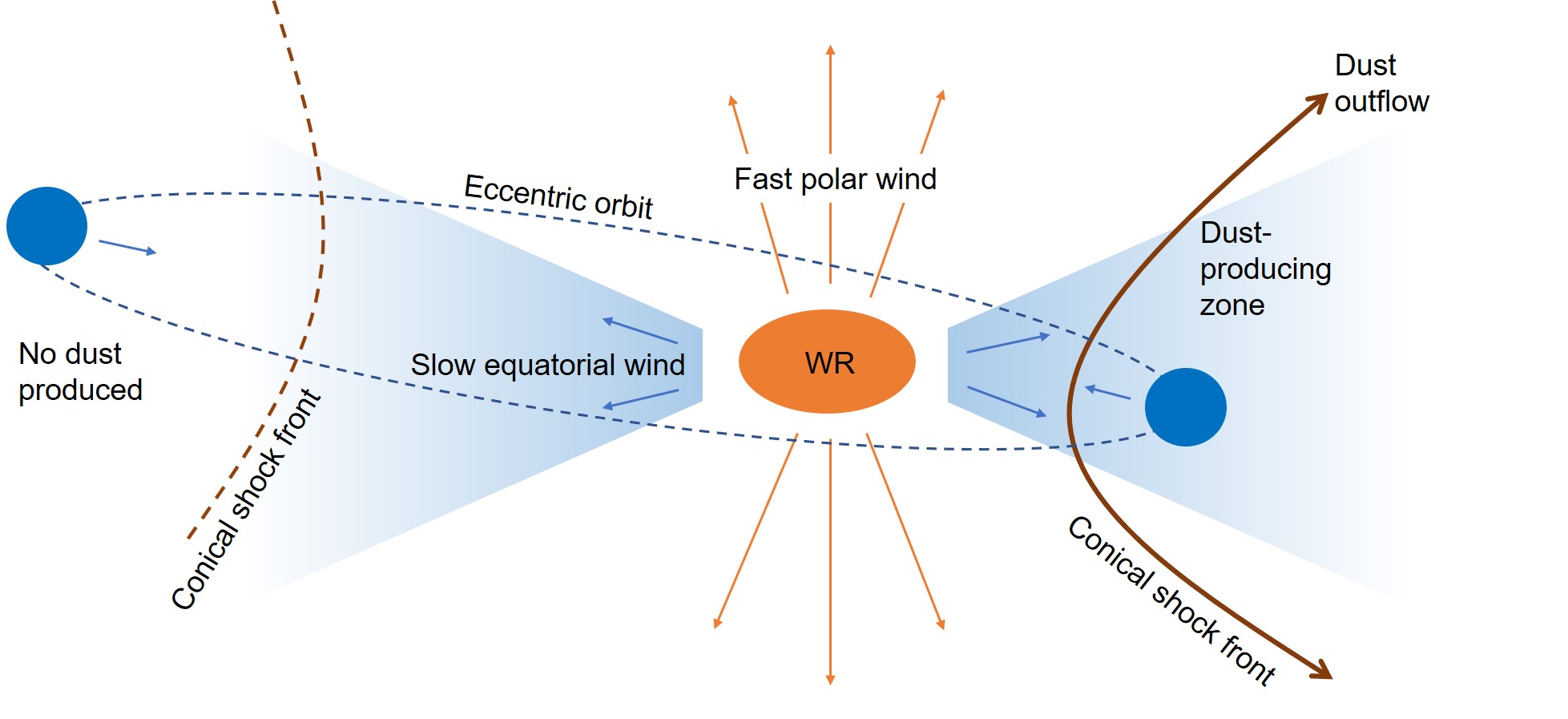}
\caption{Schematic diagram of the anisotropic wind model which postulates a fast polar wind, evident in spectroscopic data, with a slow equatorial wind responsible for the more stately expansion of the dust plume. Only sectors of the orbit around periastron result in conditions suitable for dust production. }
\label{fig:Anisotropic}
\end{figure*}

We therefore encourage ideas from the stellar wind community: this conundrum now seems shored up by firm data on the winds, the identification of the stellar components and the distances. 
So far the main contender remains the anisotropic wind model proposed by \citep{Callingham2019}, depicted in Fig~\ref{fig:Anisotropic}, where the WR binary is proposed to be capable of launching wind speeds varying across latitude. 
The central WR star drives a significantly slower wind in the equatorial direction than in the polar direction (which is probed by spectroscopy), resulting in dust formation embedded within the slower terminal wind in the equatorial plane. 

This may indicate rapid stellar rotation -- a phenomenon associated with the generation of anisotropic winds \citep{groh10,steffen14,gagnier19}. Identifying Apep as containing a rapidly-rotating WR has broader astrophysical significance. Under the collapsar model, rapidly rotating Wolf-Rayet stars are likely progenitors of long-duration gamma-ray bursts \citep{woosley93,Thompson1994,macfadyen99,Macfayden2000, Woosley2006}. The anisotropic wind model would make Apep as a potential candidate for an LGRB progenitor, the first such system to be discovered in the galaxy. 

Future work may also like to explore the alternative theory of magnetic confinement which may be capable of slowing magnetic equatorial winds \citep{UdDoula2003}.

\section{Conclusions}
\label{sec:sum}

The analysis of near-infrared NACO imagery primarily in $H$/$K$-bands revealed the structure of the dust engine embedded in the centre of Apep's Pinwheel nebula for the first time, delivering the detection of a binary with a $47\pm6$\,mas apparent separation. The binary at the centre of Apep is consistent with spectroscopic expectations \citep{Callingham2020}, and the observed position angles of $274\pm2^{\circ}$ (Apr 2016) and $278\pm3^{\circ}$ (Mar 2019) were robust against the choice of model and are in close agreement with radio observations for the colliding-wind shock front \citep{Marcote2020}. 
The change in orientation between the two epochs, although of only modest statistical significance, is in the right direction and of the right magnitude for the expected binary orbit.
Further towards the thermal-infrared (from $L$-band onwards), flux from local hot dust rose to dominate the energetics within the region of the central binary. 

Our new geometric model incorporating eccentricity offers an improved fit to the details of the dust plume compared to previously published work, constraining the orbital parameters and distance to the colliding-wind binary. The model favours an orbital period of $125\pm20$~yr, an eccentricity of $0.7\pm0.1$ and a shock cone full opening angle of $125\pm10^{\circ}$, and suggests that the binary is presently near apastron. Assuming Apep is an episodic dust producer, the best-fit orbital parameters predict that the upcoming episode of dust production will begin in $\sim$41~yr and continue for $\sim$38~yr. 

Using three epochs of mid-infrared VISIR imagery spanning a 2-yr time baseline, improved proper motion studies extracted an updated $80\pm11$\,mas\ yr$^{-1}$ expansion speed, translating to approximately $910\,^{+210}_{-290}\,$km\,s$^{-1}$ at a distance of $2.4^{+0.2}_{-0.5}$\,kpc \citep{Callingham2019} favoured by orbital constraints. Confirmation of the observational dichotomy between the slowly expanding dust plume despite spectroscopically confirmed fast radial winds \citep{Callingham2020} provides support for the anisotropic wind model which puts forward Apep as the first potential LGRB progenitor candidate to be identified in the galaxy. 

The authors encourage further observational and theoretical progress on this fascinating system. New data with the ability to reveal presently hidden structural detail (both close to the central binary and beyond the main mid-infrared dust plume), such as SPHERE, MATISSE or ALMA imaging, may be particularly useful, as is radiative transfer modelling of the plume and theoretical exploration of the colliding wind physics with anisotropic mass loss.

\section*{Acknowledgements}
This work was performed in part under contract with the Jet Propulsion Laboratory (JPL) funded by NASA through the Sagan Fellowship Program executed by the NASA Exoplanet Science Institute. 

YH, PT and AS acknowledge the traditional owners of the land, the Gadigal people of the Eora Nation, on which the University of Sydney is built and this work was carried out. JRC thanks the Nederlandse Organisatie voor Wetenschappelijk Onderzoek (NWO) for support via the Talent Programme Veni grant. BJSP acknowledges being on the traditional territory of the Lenape Nations and recognizes that Manhattan continues to be the home to many Algonkian peoples. We give blessings and thanks to the Lenape people and Lenape Nations in recognition that we are carrying out this work on their indigenous homelands. BM acknowledges support from the Spanish Ministerio de Econom\'ia y Competitividad (MINECO) under grant AYA2016-76012-C3-1-P and from the Spanish Ministerio de Ciencia e Innovaci\'on under grant PID2019-105510GB-C31.

This research made use of NASA's Astrophysics Data System; \texttt{LITpro}, developed and supported by the Jean-Marie Mariotti Centre (JMMC) \citep{Chesneau2009}; the \textsc{IPython} package \citep{ipython}; \textsc{SciPy} \citep{scipy}; \textsc{NumPy} \citep{numpy}; \textsc{matplotlib} \citep{matplotlib}; and Astropy, a community-developed core Python package for Astronomy \citep{astropy}.

\section*{Data availability}
The data underlying this article are available on the ESO Archive under observing programmes 097.C-0679(A), 097.C-0679(B), 299.C-5032(A), 0101.C-0726(A) and 0102.C-0567(A). 

\bibliographystyle{mnras}
\input{main.bbl}


\appendix

\section{Near-infrared images}
\begin{figure*}
\centering
\includegraphics[width=14cm]{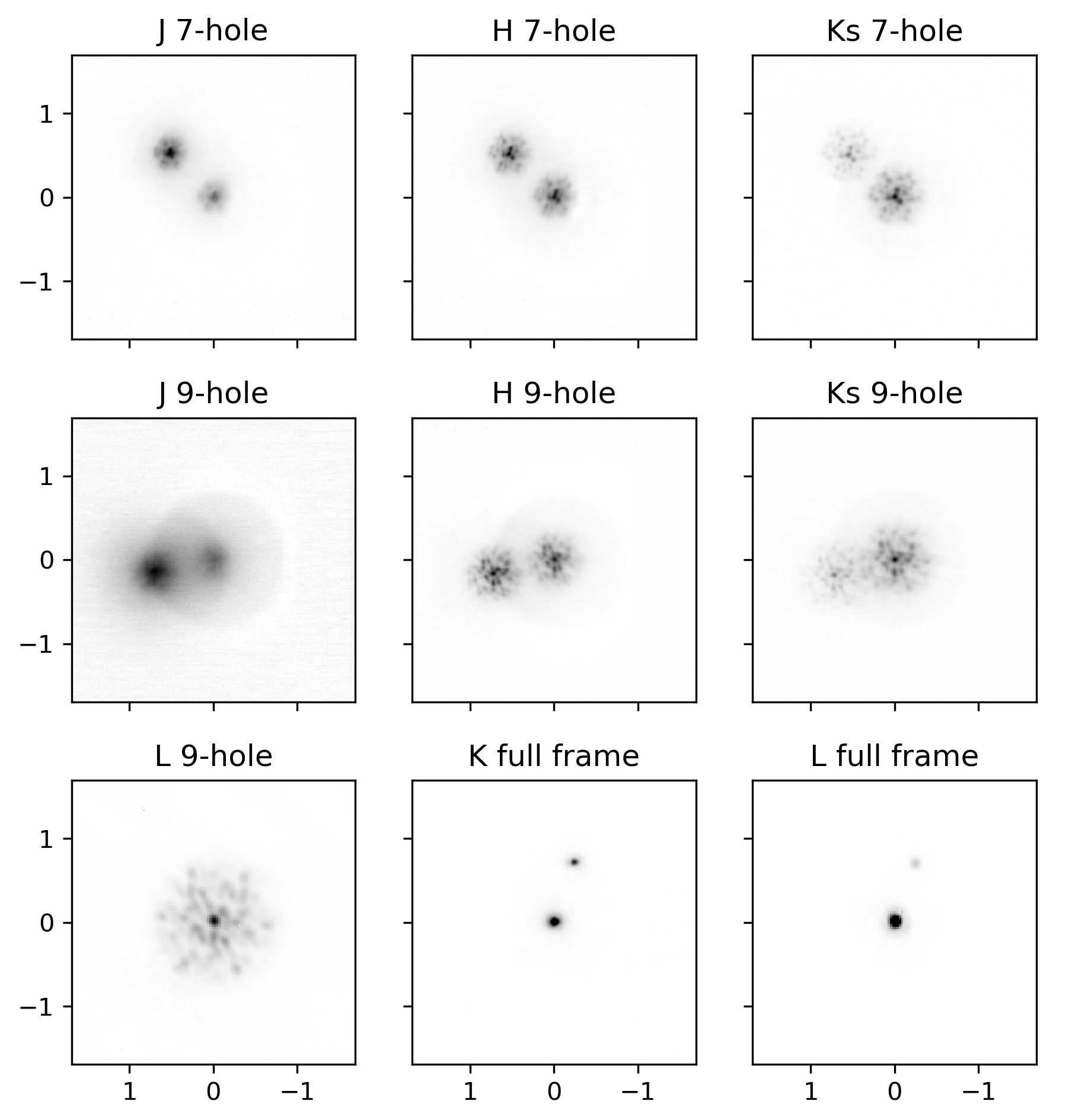}
\caption{Example cleaned data frames of Apep from each band and mask (stacked over all frames in each sample cube). Images have been scaled to a common plate scale (axes are in units of mas), however the position angle which varies with epoch and time-of-observation has not, causing the orientation of the binary to appear to change. }
\label{fig:NACO}
\end{figure*}

A series of cleaned near-infrared images of Apep observed with NACO are presented in Figure~\ref{fig:NACO}. Both the central binary (centred) and northern companion appear in each frame, the relative flux between which can be seen to vary across the spectral bands. 

\section{Edge detection}
\label{edgdet}

\subsection{Displacement visualisation}

\begin{figure*}
\centering
\includegraphics[width=15cm]{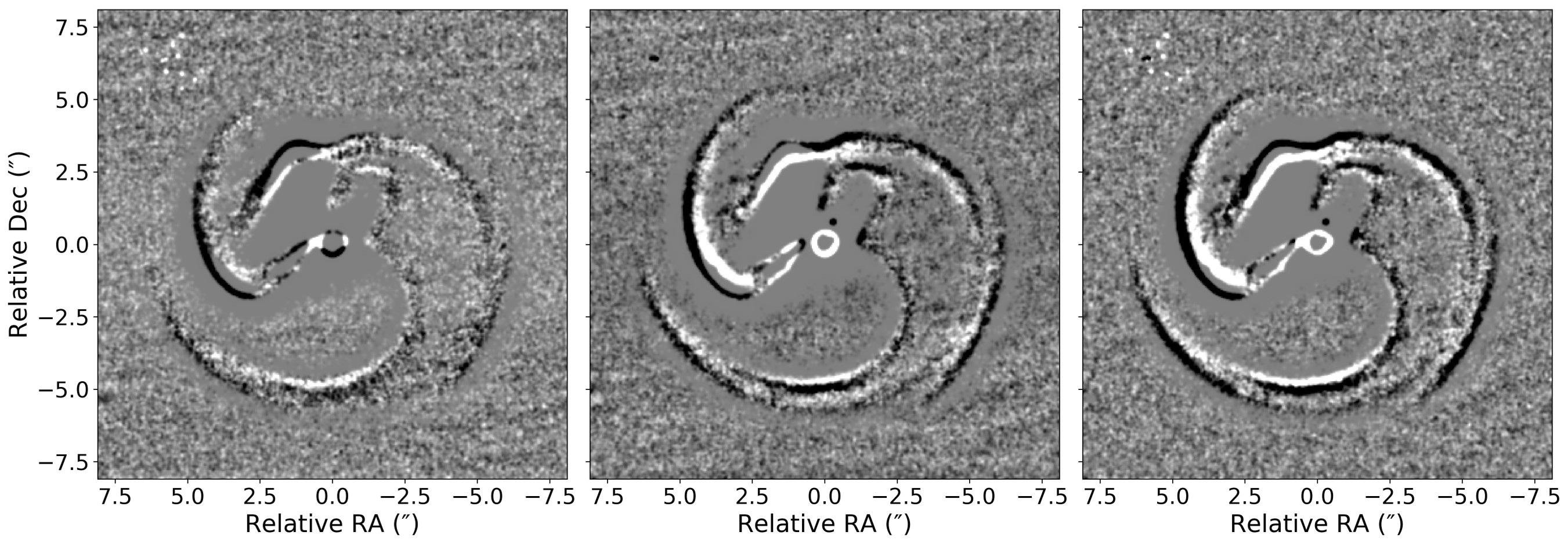}
\caption{High-pass filtered difference images in which newer epochs have been subtracted from older. Data from 2016 and 2017 (left), 2017 and 2018 (middle), and 2016 and 2018 (right). Negative images (black) always lie exterior to positive ones (white) implying the expansion of the dust plume is clearly visible across all combinations.}
\label{fig:expand1}
\end{figure*}

\begin{figure}
\centering
\includegraphics[width=7cm]{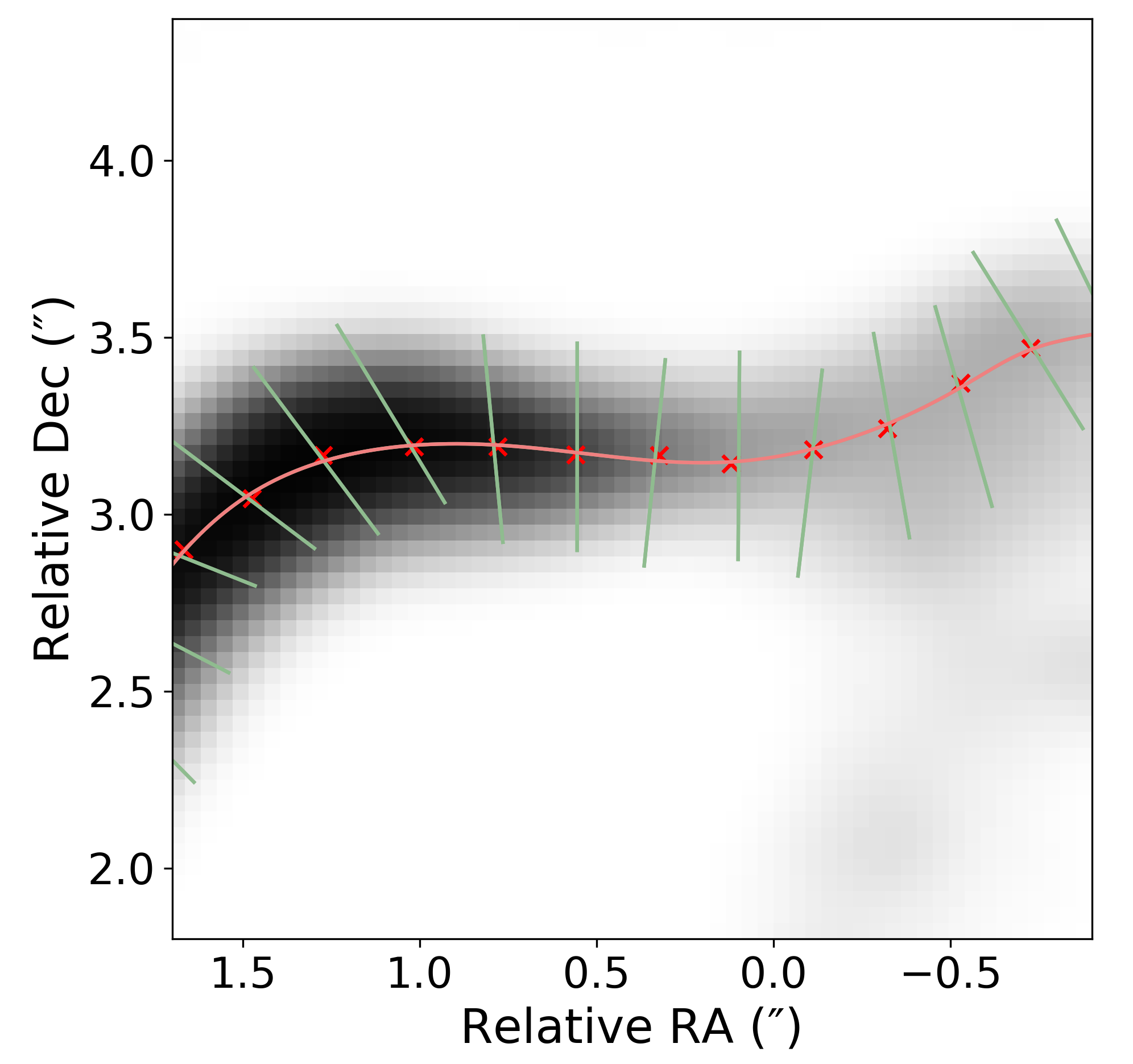}
\caption{A zoomed in example of \texttt{ridge-crawler} in action showing sample points, search slices and a fitted spline. }
\label{fig:rc}
\end{figure}

In order to register displacements of image components over time, specific features or structures must be identified and tracked. 
High-pass filtering is a useful tool in this context, enhancing features such as edges, essentially revealing the skeleton of Apep's dust. Such underlying structural elements of the plume, illustrated in Fig~\ref{fig:rc}, can be accurately registered at each epoch. Even without further processing, the expansion of the plume may be directly confirmed in a simple difference image between epochs as given in Fig~\ref{fig:expand1}.
This reveals that the edges of the spiral plume from the older epoch of all pairs are always completely contained within those of the newer epoch, implying a consistent outwards motion of the plume over time. 
Although such confirmation is helpful, it is difficult to extract robust estimates of the rate of expansion from such difference images; an issue that motivated the more advanced procedures discussed below.

\subsection{Displacement extraction algorithm}
Given the small dust displacements of Apep's plume of order one pixel per year, we implemented a custom algorithm, \texttt{ridge-crawler} to locate edges at the sub-pixel level. 

The algorithm is initialised with a user-specified location and direction, from which the it begins taking discrete steps along the crest of the ridge in the image data. At each iteration, \texttt{ridge-crawler} looks $n$ pixels ahead in the direction interpolated from its previous path, takes the profile of a slice of pixels in the orthogonal direction, and fits a parabola through the three brightest pixels along that slice. \texttt{ridge-crawler} then calculates the location of the peak of the parabola, records it as the next sample along the ridge, and moves to this location ready for a new iteration. The action of this algorithm is depicted in Fig~\ref{fig:rc}. 

The algorithm fits spline functions to the sample points of each ridge and densely samples the splines. It then projects rays from the centre of the image, intercepting the splines at both epochs, calculating the radial displacement in polar coordinates across the range of shared angles between each corresponding pair of splines. Sample code is available at \href{https://github.com/yinuohan/Apep/blob/master/VISIR/ridge_crawler.py}{github.com/yinuohan/Apep}. 

We note that the errors associated with the extracted dust expansion speed are not statistically independent. Both the time-varying PSF and the \texttt{ridge-crawler} algorithm introduce correlated errors into the displacement measurements for each point along a given ridge. To estimate the error of the mean displacement of each ridge, we calculated an adjusted SEM using the ``true'' number of independent sample points (adjusted for the autocorrelation distance of each ridge).

\subsection{Deriving expansion speed from ridge displacements}
\label{sec:modelspeed}
The two-dimensional geometry of the dust plume arises from the projection onto the sky plane of an optically thin three-dimensional structure. 
Using the geometric model for the plume, we calibrated the apparent expansion speed of individual ridges against the physical outflow speed of the system. Since the \texttt{ridge-crawler} algorithm introduces correlated error (newer sample points along each ridge are not independent from previous ones), multiple independent measurements were conducted. Four pairs of images of Apep's dust plume were generated using the geometric model separated by 646 days pairwise (consistent with the time between the 2016 and 2018 VISIR epochs) and a 1~yr offset between different pairs. Each edge-like structure identified in the mid-infrared images was also identified in the model images and sampled with \texttt{ridge-crawler} pairwise, the displacement of which were averaged across pairs. The results are presented in Table~\ref{modelspeed}. 

\begin{table}
\begin{center}
\caption{Normalised mean expansion speed of ridges $a$ to $f$ detected by \texttt{ridge-crawler} averaged across four sets of images generated by the geometric model. Measured speeds were normalised against the wind speed of 80 mas\ yr$^{-1}$ (the mean expansion speed across all ridges between the 2016 and 2018 VISIR epochs) that was injected into the model. Uncertainties were estimated using the standard error of the mean. }
\label{modelspeed}
\begin{tabular}{ c c } 
 \hline
 Ridge name & Normalised expansion speed\\
 \hline
 \textit{ a} & 0.98 $\pm$ 0.07 \\
 \textit{ b} & 0.81 $\pm$ 0.09 \\
 \textit{ c} & 0.98 $\pm$ 0.08 \\
 \textit{ d} & 1.03 $\pm$ 0.08 \\
 \textit{ e} & 0.85 $\pm$ 0.08 \\
 \textit{ f} & 0.97 $\pm$ 0.08 \\
 \hline
 \textbf{Mean} & \textbf{0.94} $\pm$ \textbf{0.08}\\
 \hline
\end{tabular}
\end{center}
\end{table}

The results suggest that expansion speed of ridges $a$, $c$, $d$ and $f$ are consistent with the true expansion rate of the system, and therefore these features arise from limb-brightened edges that lie in or close to the plane of the sky. 
On the other hand, ridges $b$ and $e$ are significantly slower, implying that dust producing these likely has some projection to the line of sight. 
We exploited this finding to inform the ridges chosen to estimate the physical expansion speed of the dust plume, as described in Section~\ref{sec:VISIR}. 

\section{Geometric model}
\label{geomod}

\subsection{Model construction}
We adopt a geometric model based upon the Pinwheel mechanism, where we assume that the production of dust occurs on the surface of a conical shock -- the downstream shape that results from the collision of the two winds (the curved region at the nose of the shock is too small to be seen here). 
This is modelled by forming circular rings at the wind-wind stagnation point, which are inflated and carried linearly away from the geometric centre of the system at the terminal wind speed ($w$). The geometry of the wind-wind interface carrying the dust is specified by the cone opening angle ($\theta_w$), a physical parameter determined by the wind momentum ratio. The present position angle of the binary is defined by a time offset ($T_0$) from periastron. The simulated orbit begins a specified number of orbital periods ($n_\textsubscript{circ}$) ago in which dust production occurs over a specified range of orbital phase. The binary star evolves with an orbital period ($P$) and eccentricity ($e$), and the system is further specified by additional binary-star orbital elements associated largely with viewing angle, the position angle of the ascending node ($\Omega$), argument of periastron ($\omega$) and the inclination of the orbital plane ($i$), used to project the model into a final image. In essence, the model effectively creates a conical surface wrapped into a spiral by the orbit of the binary and viewed at user-specified angles in three-dimensional space. 

\subsection{Numerical optimisation approaches}
Model optimisation with a numerical algorithm such as the Markov Chain Monte Carlo (MCMC) was attempted, but we were unable to find conditions where the code would converge. The greatest hindrance to this approach is the lack of an acceptable metric that can quantify correspondence between ridges in the image and model at arbitrary clocking angles, a problem made worse by the richness of structures that such a model can generate and the high-dimensional search space. Indeed, since it is the geometry and location of the spatial features that the model is attempting to create, pixel values in isolation carry little structural information when simply applying a $\chi^2$ difference metric between the data and model-generated images. 

In an attempt to address this challenge, we developed an algorithm to convert both the high-pass filtered data and model-generated images into binary pixel values, effectively turning them into ``pencil sketches'', which was designed to preserve only the main skeleton of the plume. The pencil sketches were subsequently Gaussian-blurred (to provide smooth image gradients) before computing the $\chi^2$ difference between the data and model, which can be fed into a numerical optimisation algorithm. However, the structural outlines generated by the model remained many and complex and, together with the high-dimensional nature of the search space, the MCMC was unable to converge. Following these attempts, we instead manually explored the parameter space informed by our knowledge of the physical parameters of the binary.

\subsection{Visualisation}
\begin{figure*}
\centering
\includegraphics[width=16cm]{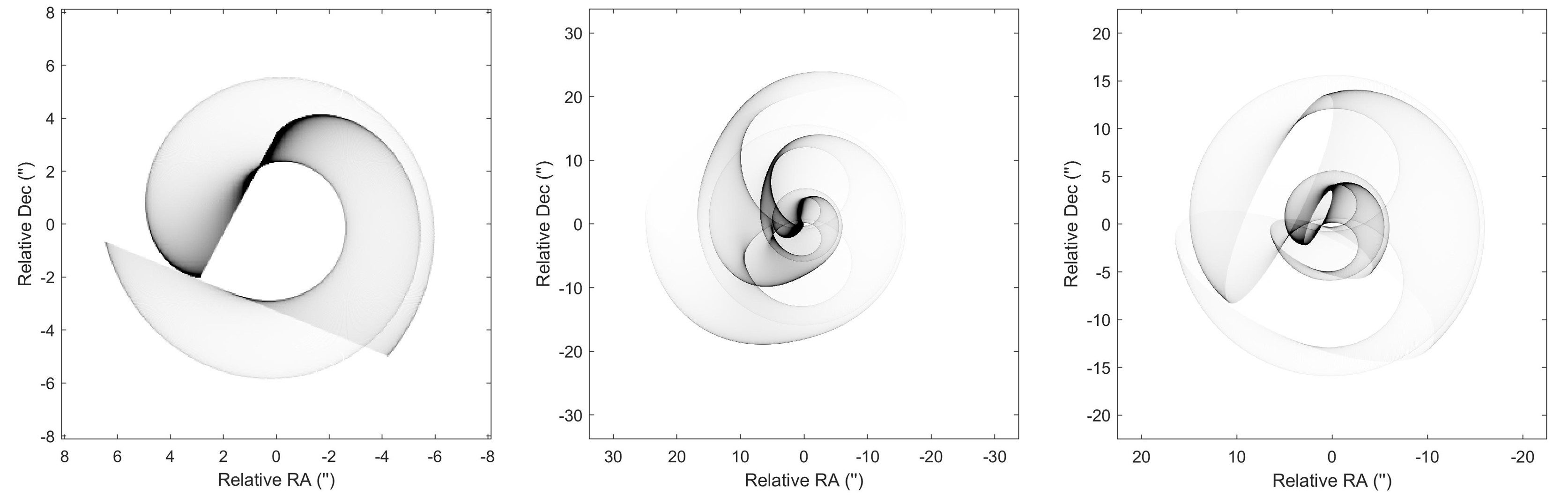}
\caption{Left: pole-on view of the dust plume (inclination = 0$^{\circ}$). Middle: the geometry of the dust plume assuming constant dust production over 2.5 periods (without the threshold effect) Right: the location and geometry of the dust plume produced from a prior cycle predicted by the threshold effect are presented, overlaid on the model of the dust plume visible in the mid-infrared.  }
\label{fig:modelpics}
\end{figure*}

Figure~\ref{fig:modelpics} provides additional images to assist the interpretation of the geometric model and its implications. A pole-on view of the dust plume, the hypothetical geometry of the dust plume without the threshold effect and the location and geometry of a more distant dust cycle predicted by the threshold effect are presented. An additional animated figure showing the formation of the dust plume over time is available as an online supplementary material. 

\bsp	
\label{lastpage}
\end{document}